\documentclass[journal]{IEEEtran}

\usepackage{bbm}
\usepackage{caption}
\usepackage{subfigure}
\usepackage{graphicx}
\usepackage{latexsym}
\usepackage{diagbox}
\usepackage{changepage}
\usepackage[fleqn]{amsmath}
\usepackage{amsmath}
\usepackage{amsfonts}
\usepackage{indentfirst}
\usepackage{CJK}

\usepackage{indentfirst}
\usepackage[varg]{txfonts}
\usepackage{stfloats}
\usepackage{multirow}%
\usepackage{booktabs}
\usepackage{color,soul}
\usepackage{epstopdf}
\usepackage{float}
\usepackage{bm}
\usepackage{makecell}
\usepackage{array}
\usepackage{bm}
\usepackage{mathtools}
\usepackage{geometry}

\usepackage[linesnumbered,ruled,commentsnumbered,longend]{algorithm2e}
\usepackage{cuted}
\makeatletter

\newcommand{\Rmnum}[1]{\expandafter\@slowromancap\romannumeral #1@}

\makeatother
\makeatletter

\newcommand{\qed}{\nobreak \ifvmode \relax \else
	\ifdim\lastskip<1.5em \hskip-\lastskip
	\hskip1.5em plus0em minus0.5em \fi \nobreak
	\vrule height0.75em width0.5em depth0.25em\fi}

\SetAlgoLongEnd

\begin{document}
\title{{Wideband Beamforming for STAR-RIS-assisted THz Communications with Three-Side Beam Split}}

\author{Wencai Yan, Wanming Hao,~\IEEEmembership{Senior Member,~IEEE,} Gangcan Sun, Chongwen Huang,~\IEEEmembership{Member,~IEEE,}\\ Qingqing Wu,~\IEEEmembership{Senior Member,~IEEE}
	
	\thanks{W. Yan, W. Hao and G. Sun are with the School of Electrical and Information Engineering, Zhengzhou University, Zhengzhou 450001, China. (E-mail: yanwencai001@163.com, \{iewmhao, iegcsun\}@zzu.edu.cn).}
	\thanks{C. Huang is with College of Information Science and Electronic Engineering, Zhejiang University, Hangzhou 310027, China. (E-mail: chongwenhuang@zju.edu.cn).}
    \thanks{Q. Wu is with the Department of Electronic Engineering, Shanghai Jiao Tong University, Shanghai 201210, China. (E-mail:
qingqingwu@sjtu.edu.cn).}}

\maketitle

\begin{abstract}
In this paper, we consider the simultaneously transmitting and reflecting reconfigurable intelligent surface (STAR-RIS)-assisted THz communications with three-side beam split. Except for the beam split at the base station (BS), we analyze the double-side beam split at the STAR-RIS for the first time.  To relieve the double-side beam split effect, we propose a time delayer (TD)-based fully-connected structure at the STAR-RIS. As a further advance, a low-hardware complexity and low-power consumption sub-connected structure is developed, where multiple STAR-RIS elements share one TD. Meanwhile, considering the practical scenario, we investigate a multi-STAR-RIS and multi-user communication system, and a sum rate maximization problem is formulated by jointly optimizing the hybrid analog/digital beamforming, time delays at the BS as well as the double-layer phase-shift coefficients, time delays and amplitude coefficients at the STAR-RISs. Based on this, we first allocate users for each STAR-RIS, and then derive the analog beamforming, time delays at the BS, and the double-layer phase-shift coefficients, time delays at each STAR-RIS. Next, we develop an alternative optimization algorithm to calculate the digital beamforming at the BS and amplitude coefficients at the STAR-RISs. Finally, the numerical results verify the effectiveness of the proposed schemes.
\end{abstract}


\IEEEpeerreviewmaketitle

\section{Introduction}
Terahertz (THz) has attracted great attention in future 6G networks due to its ultra-wide bandwidth~\cite{ref1},~\cite{ref2}. However, the ultra-high frequency of THz signals results in the serious propagation loss, limiting the transmission distance~\cite{ref3}.  Fortunately, massive multiple-input multiple-output (MIMO) is usually exploited to realize high-gain directional beams, which can effectively improve the coverage range of THz communications. Additionally, the traditional fully-digital antenna structure will bring huge power consumption and hardware complexity for THz massive MIMO communications. Therefore, the hybrid analog/digital antenna structure are usually applied~\cite{ref4},~\cite{ref5}.

Furthermore, THz communications exhibit the characteristic of poor diffraction, rendering it susceptible to signal blockages, and thus breaks the line-of-sight (LoS) link. As such, reconfigurable intelligent surface (RIS)~\cite{ref6},~\cite{ref6_0} can be implemented to tackle this limitation, which is equipped with massive low-cost passive components. However, RIS is limited to reflecting incident wireless signals, requiring both the transmitter and receiver to be deployed on the same side of the RIS. To extend the scope of the coverage comprehensively, an innovative approach known as simultaneously transmitting and reflecting reconfigurable intelligent surfaces (STAR-RIS) is introduced~\cite{ref7}-\cite{refB}.  STAR-RIS is able to concurrently transmit and reflect incident signals on both sides of the surface~\cite{ref9}. Moreover, each individual element of STAR-RIS applies two distinct and adjustable phase-shift values, catering to both the transmitted and reflected~signals.

Therefore, it is promising to apply the STAR-RIS to THz communications. Whereas, due to the frequency-independent phase shifter (PS) at the BS and frequency-independent element at the STAR-RIS, there will exist beam split~\cite{ref10},~\cite{ref11}, seriously  effecting the system performance. Consequently, in this paper, we mainly analyze the beam split and design the wideband beamforming in the STAR-RIS-assisted THz communications.

\subsection{Related Works}
Several works have predominantly focused on addressing the beam split problem at the BS~\cite{ref12}-\cite{ref16}. A fundamental solution is to replace all PSs with time delayer (TD)~\cite{ref13}. Unlike PS which produces a fixed phase across the entire bandwidth, the TD can provide a controllable time delay, inducing the different phase across overall frequencies. This frequency-dependent characteristic of TD effectively mitigates the beam split problem. However, replacing all PSs with TD is impractical, owing to the associated huge hardware costs within the THz band. Instead, an alternative approach proposed by~\cite{ref14}-\cite{ref16} is to introduce a limited number of TDs between radio frequency (RF) chain and PSs so as to construct a TD-aided hybrid beamforming architecture. Specifically, in~\cite{ref14}, authors introduced an energy-efficient fixed TD structure that deploys low-resolution discrete time delays. Although this scheme can effectively reduce the power consumption, it is unable to fully solve the beam split issue. In~\cite{ref15}, authors presented a novel THzPrism architecture where TDs are arranged serially. Nevertheless, the hardware complexity of this architecture is high due to huge number of TDs. Furthermore, an effective delay-phase scheme was proposed in \cite{ref16} to eliminate the beam split effect, where a few TDs insert between PSs and RF chains in parallel.

Recently, several works have considered the beam split problem at the RIS and investigated the wideband beamforming~\cite{ref17}, \cite{ref18}. Particularly, the association between the deployment of RIS and the beam split effect was analyzed in~\cite{ref18}. It is noteworthy that the methods employed in~\cite{ref17},~\cite{ref18} still depend on the frequency-independent beamforming architectures, which is unable to solve the beam split of the RIS essentially. Taking a significant step forward, the authors in~\cite{ref19} analyzed the beam split effect based on far-field and near-field in THz RIS-assisted communications. To overcome the beam split effect, they introduced delay-adjustable metasurfaces at the RIS. Nevertheless, implementing delay-adjustable metasurfaces at each RIS element may be impractical due to the associated complexity. As a result, an element-grouping strategy of RIS was proposed in \cite{ref20},~\cite{ref21}, and each sub-surface was equipped with a TD, which is capable of effectively mitigating the beam split effect. However, it is worth noting that~\cite{ref19},~\cite{ref20} only considered the beam split at the RIS, while the beam split at the BS was not investigated. In~\cite{ref21}, although the authors considered the beam split at both the BS and RIS, it is limited to the single-user scenario.

With the analysis above, there is no relevant study to jointly consider the beam split effect and beamforming optimization at the THz BS and STAR-RIS, which is~challenging.
\subsection{Main Contributions}
To mitigate the three-side beam split, we apply the limited number of TDs to both the BS and STAR-RIS. Then, the joint beamforming optimization problem is considered. The main contributions are summarized as follows:
\begin{itemize}
\item[$\bullet$]
We first analyze the double-side beam split effect at the STAR-RIS, and then propose a TD-aided fully-connected structure, where each STAR-RIS element adopts one PS and one TD to eliminate the beam split.  As a further advance, we introduce a low-cost sub-connected structure, where each STAR-RIS element with double-layer PSs, and multiple STAR-RIS elements are divided into a sub-surface to share one common TD.
\end{itemize}
\begin{itemize}
\item[$\bullet$]
Next, we investigate the distributed STAR-RISs-assisted THz communications, where the TD-based hybrid analog/digital structure at the BS and the TD-based sub-connected structure at each SATR-RIS are considered, respectively. Based on this, a sum
rate maximization problem is formulated by jointly optimizing the hybrid analog/digital beamforming, time delays at the BS as well as the double-layer phase-shift coefficients, time delays and amplitude coefficients at the STAR-RISs. To tackle this joint optimization problem, we first divide group for each STAR-RIS, and then derive the analog beamforming, time delays at the BS, and the double-layer phase-shift coefficients, time delays at each STAR-RIS. Finally, the original problem is transformed to a simplified problem to maximize the sum rate.

\end{itemize}
\begin{itemize}
\item[$\bullet$]
To solve the above transformed problem, an alternative iterative algorithm is proposed. Specifically, the digital beamforming at the BSs and the amplitude coefficients at the STAR-RISs are first decoupled by lagrangian dual reformulation (LDR) and multidimensional complex quadratic transform (MCQT). Next, by alternatively tackling the separated quadratically constrained quadratic program (QCQP) subproblems, the final solutions are obtained. In addition, we also provide the analysis with the robustness of the joint wideband beamforming framework versus the impacts of imperfect channel state information (CSI).

\end{itemize}

\textit{Notations:}  Lower-case, bold-face lowercase and bold-face uppercase boldface represent as scalar, vector and matrix, respectively. $(\cdot)^{{T}}$ indicates the transpose. $(\cdot)^{{H}}$ indicates the Hermitian transpose. $\rm diag(\cdot)$ is a diagonal matrix. $\otimes$ represents the Kronecker product.  $\operatorname{Tr}\{\cdot \}$ means the trace of its argument. $|\cdot|$ is the modulus of a complex number. $\left\|\cdot \right\|$ represents the Frobenius norm  of matrix. $\ln(\cdot)$ denotes natural logarithm. $\mathbf{I}_{N}$ denotes the $N \times N$ identity matrix.  $\mathbb{C}^{x \times y}$ is the space of $x \times y$ complex matrix.  $\rm {Re}(\cdot)$ represents the real component of a complex number. $\mathbb{E} \{\cdot \}$ is the expectation operator.
\section{Beam Split Effect Analysis and STAR-RIS Structure Design}
In this section, the double-side beam split at the STAR-RIS is first analyzed. Then, we employ a TD-based fully-connected structure at the STAR-RIS to mitigate the double-side beam split effect. Meanwhile, a low-cost sub-connected structure is developed.
\subsection{Beam Split Effect Analysis}
As shown in Fig. 1, there are three-side beam split in the STAR-RIS-assisted THz communications system. Actually, there exist several studies investigating the beam split effect at the BS. Thus, we emit this analysis, and more details can refer to~\cite{ref10},~\cite{ref14}. Next, we mainly focus on analyzing the double-side beam split at the STAR-RIS. Let $\mathbb{R}$ and $\mathbb{T}$ represent reflection and transmission side, respectively. To simplify the expression, we assume that the BS owns single antenna and the STAR-RIS consists of $N_{\rm RIS}=N_1 N_2$ elements, where $N_1$ and $N_2$ are the number of rows and columns, respectively. In this paper,  it is assumed that the direct links between the BS and users are blocked by obstacles. Meanwhile, the perfect CSIs of all links are assumed to be available by existing advanced channel acquisition methods~\cite{ref22}-\cite{refD}. The orthogonal frequency division multiplexing (OFDM) technique with $M$ subcarriers is adopted to address the frequency selective fading~\cite{ref24}. Denote $B$ as the bandwidth and $f_{c}$ as the central frequency, and thus the frequency at the $m$-th subcarrier is $f_{m}=f_{c}+\frac{B}{M}\left(m-1-\frac{M-1}{2}\right), m=1,2,\cdots, M$.
\begin{figure}[t]
\centering
    \label{RIS_subcarrier} 
    \includegraphics[width=8cm,height=4cm]{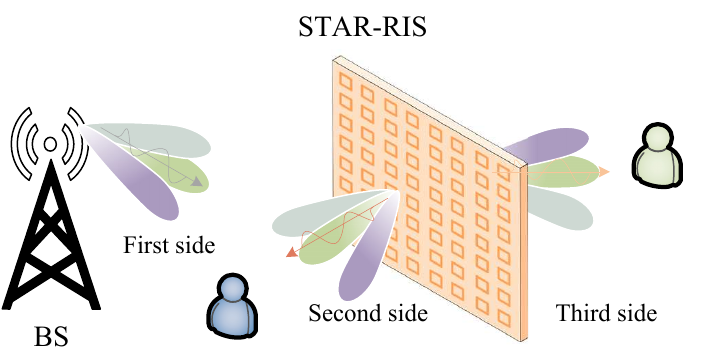}
	
	\caption{STAR-RIS-assisted THz communications with three-side beam split.}
\end{figure}

Generally, the communication channel consists of a LoS path and a few non-LoS (NLoS) paths~\cite{ref25}. However, due to the property of THz signals, the LoS path is usually dominated, hence we assume that only LoS path exists here~\cite{ref3},~\cite{ref21}. Thus, the channel matrix $\mathbf{G}_{m}$ between the BS and STAR-RIS is given by
\begin{eqnarray}
\mathbf{G}_{m}= \alpha_{1} e^{-j 2 \pi t_{1} f_{m}} \mathbf{b}\left(f_m, u_{1}, v_{1}\right),
\end{eqnarray}
where $t_{1}$ and $\alpha_{1}$ are the path delay and gain from the BS to the STAR-RIS, respectively. Specifically, the path gain is
\begin{eqnarray}
\alpha_{1}\left(f_m, d_{1}\right)=\frac{\mathrm{c}}{4 \pi f_m d_{1}} e^{-\frac{1}{2} \kappa_{a b s}\left(f_m\right) d_{1}},
\end{eqnarray}
where $c$, $\kappa_{a b s}(f_m)$ and $d_{1}$ denote the speed of light, medium absorption factor and distance from the BS to STAR-RIS, respectively. Besides, $\mathbf{b}\left(f_m, u_{1}, v_{1}\right)$ represents the steering vector of the STAR-RIS, which is written as~\cite{ref26}
\begin{eqnarray}
\begin{split}
\!\!\mathbf{b}\left(f_m, u_{1}, v_{1}\right)
&=\frac{1}{\sqrt{N_{\rm{RIS}}}}[1, \ldots, e^{j  \pi \xi_{m}(n_{1} \sin u_{1} \sin v_{1}+n_{2} \cos v_{1})}, \\
&\ldots, e^{j  \pi  \xi_{m}((N_{1}-1) \sin  u_{1} \sin v_{1}+(N_{2}-1) \cos v_{1})}]^{T},
\end{split}
\end{eqnarray}
where $\xi_{m}=\frac{f_m}{f_c}$ is the relative frequency, $u_{1}\in[-\pi / 2, \pi / 2]$ and $v_{1} \in[0, \pi]$ are the azimuth and elevation angles of arrival (AOA) of the STAR-RIS, respectively.

Similarly, the channel vector $\mathbf{h}_{m,i}$ between the STAR-RIS and reflection/transmission user is expressed as
\begin{eqnarray}
\mathbf{h}_{m,i}=\alpha_{i} e^{-j 2 \pi t_{i} f_{m}} \mathbf{b}\left(f_m, u_i, v_i\right)^{T}, \forall i \in \{ \mathbb{R}, \mathbb{T}\},
\end{eqnarray}
where $t_{i}$ and $\alpha_{i}$ represent the path delay and gain from the STAR-RIS to reflection/transmission user, respectively. The path gain $\alpha_{i}$ is modeled as
\begin{eqnarray}
\alpha_{i}\left(f_m, d_{i}\right)=\frac{\mathrm{c}}{4 \pi f_m d_{i}} e^{-\frac{1}{2} \kappa_{a b s}\left(f_m\right) d_{i}},
\end{eqnarray}
where $d_{i}$ is the distance. $u_{i}\in[-\pi / 2, \pi / 2]$ and $v_{i} \in[0, \pi]$ denote the azimuth and elevation angle of departure (AoD) at the STAR-RIS, respectively.

Based on the system model, the reflection/transmission normalized array gain $\mathrm{g}_{\mathrm{con},i}\left(f_m\right)$, $\forall i \in \{ \mathbb{R}, \mathbb{T}\}$ of the STAR-RIS with conventional structure on the $m$-th subcarrier is computed as
\begin{eqnarray}
\begin{aligned}
& \mathrm{g}_{\mathrm{con},i}\left(f_m\right)=\left|\mathbf{b}\left(f_m, u_{i}, v_{i}\right)^{T} \mathbf{\Phi}_i \mathbf{b}\left(f_m, u_{1}, v_{1}\right)\right| \\ & =\frac{1}{N_{\mathrm{RIS}}}\left|\sum_{\mathrm{n}_1=0}^{N_1-1} \sum_{\mathrm{n}_2=0}^{N_2-1} e^{j \pi \xi_{m} \left[n_1 (\varsigma_i+\varsigma_1)+n_2(\eta_i+\eta_1) \right]} \cdot e^{j \phi_{i,n}}\right|,
\end{aligned}
\end{eqnarray}
where $n=N_2 n_1+n_2+1$, $\varsigma_i=\sin u_i \sin v_i$, $\varsigma_1=\sin u_1 \sin v_1$, $\eta_i=\cos u_i$, $\eta_1=\cos u_1$. Besides, $\mathbf{\Phi}_i=\operatorname{diag}\left(\varphi_{i,1}, \cdots, \varphi_{i,N_{\rm{RIS}}}\right)$ denotes the reflection/transmission coefficient of the STAR-RIS with $\varphi_{i,n}= \beta_{i,n} e^{j \phi_{i,n}}$, where $\beta_{i,n} \in[0,1]$ and $\phi_{i,n} \in[0, 2 \pi)$ represent the reflection/transmission amplitude and~phase, respectively. Moreover, we assume that the reflection/transmission amplitude $\beta_{i,n}$ takes an arbitrary constant in this subsection, while satisfying the energy conservation constraint.
\begin{figure}[t]
\centering
    \label{RIS_subcarrier} 
    \includegraphics[width=8.5cm,height=5.5cm]{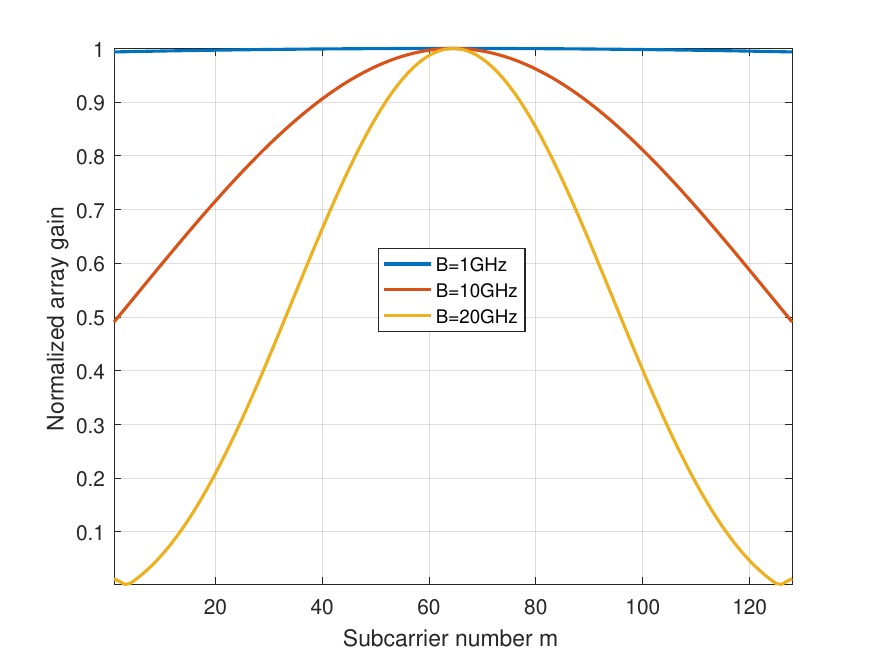}
	
	\caption{\!Normalized array gain under different bandwidths.}
\end{figure}
To generate a beam at target direction, we formulate the reflection/transmission phase-shift of the STAR-RIS on the basis of the center frequency $f_c$, which can be obtained by
\begin{eqnarray}
\phi_{i,n}=-\pi\left[n_1 (\varsigma_i+\varsigma_1) +n_2 (\eta_i+\eta_1)\right].
\end{eqnarray}
It can be found that the generated beams can toward to the target direction at the center frequency $f_c$. However, at other frequency $f_m$ with $\xi_{m}\neq 1$, the generated beams deviate from the target direction, which causes significant array gain loss as the normalized array gain $\mathrm{g}_{\mathrm{con},i}\left(f_m\right)$ fails to reach its maximum value at most frequencies. Thus, we can obtain
\begin{equation}
\begin{split}
& \mathrm{g}_{\mathrm{con},i}\left(f_m\right)=\frac{1}{N_{\mathrm{RIS}}}\left|\sum_{\mathrm{n}_1=0}^{N_1-1} \sum_{\mathrm{n}_2=0}^{N_2-1} e^{j \pi (\xi_{m}-1) \left[n_1 (\varsigma_i+\varsigma_1) +n_2 (\eta_i+\eta_1)\right]}\right| \\
& =\frac{1}{N_{\mathrm{RIS}}}\left|\Xi_{N_1}\left(\left(\xi_m-1\right) (\varsigma_i+\varsigma_1)\right) \Xi_{N_2}\left(\left(\xi_m-1\right) (\eta_i+\eta_1)\right)\right|,
\end{split}
\end{equation}
where $\Sigma_{n=0}^{N-1} e^{j n \pi x}=\frac{\sin \frac{N \pi}{2} x}{\sin \frac{\pi}{2} x} e^{-j \frac{(N-1) \pi}{2} x}$ and $\Xi_{N}\left(x\right)=\frac{\sin \frac{N \pi}{2} x}{\sin \frac{\pi}{2} x}$.

To further illustrate the beam split effect, we provide the normalized array gain under different bandwidths in Fig. 2. We set $f_{c}=100$ GHz, $M=128$ and $N_{\rm RIS}=256$. It can be found that the array gain loss increases with the bandwidth.  As a consequence, the benefits of utilizing the STAR-RIS and wideband THz resources are offset by beam split effect.
\begin{figure}[t]
	\centering
	\subfigure[]{
		\label{Iteration} 
		\includegraphics[width=8.5cm,height=3.5cm]{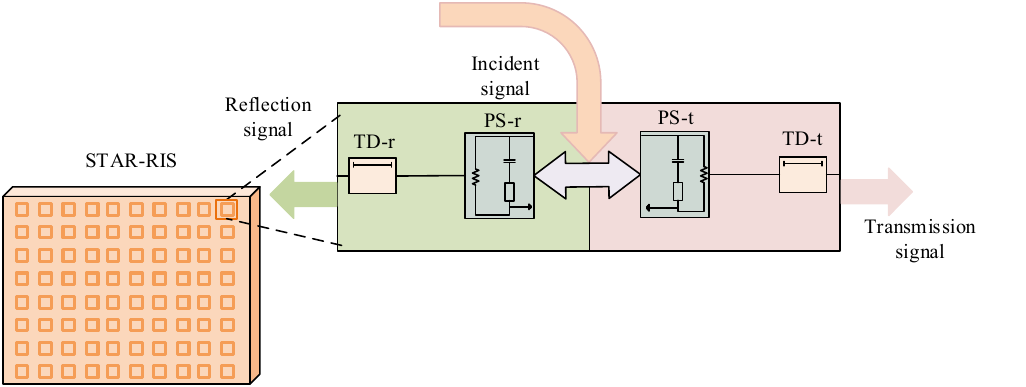}}
	\subfigure[]{
		\label{complexity} 
		\includegraphics[width=8.5cm,height=3.5cm]{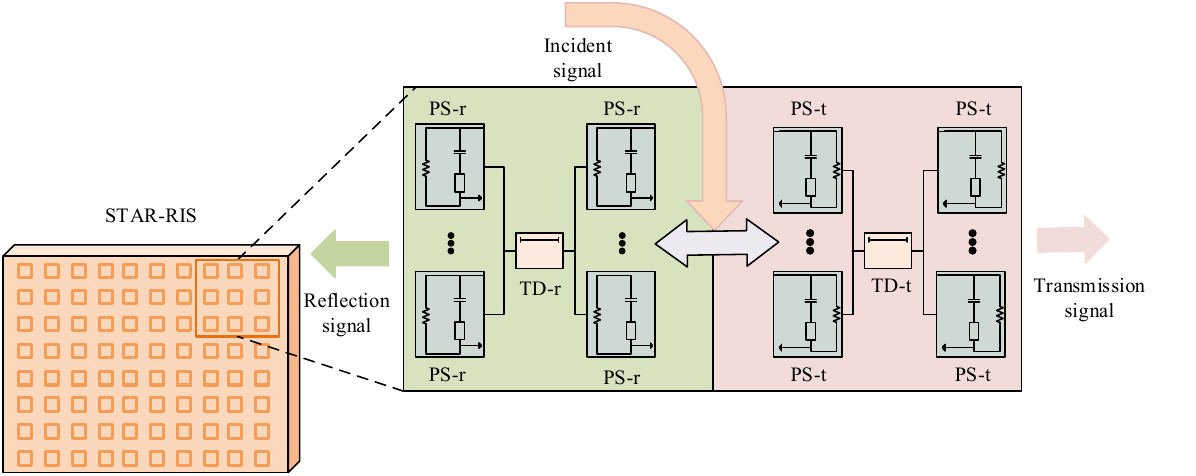}}
	\caption{(a) STAR-RIS with fully-connected structure, (b) STAR-RIS with sub-connected structure. }
\end{figure}
\subsection{STAR-RIS Structure Design}
To overcome the double-side beam split effect of the STAR-RIS, we first propose a TD-based fully-connected structure as presented in Fig. 3(a), where each of STAR-RIS element is introduced with a reflection/transmission PS and a reflection/transmission TD, ie., PS-r/PS-t and TD-r/TD-t. Thus, the reflection/transmission normalized array gain $\mathrm{g}_{\mathrm{fully},i}(f_m), \forall i \in \{ \mathbb{R}, \mathbb{T}\}$ of the STAR-RIS with fully-connected structure on the $m$-th subcarrier is computed by
\begin{equation}
\begin{split}
& \mathrm{g}_{\mathrm{fully},i}\left(f_m\right)=\left|\mathbf{b}\left(f_m, u_{i}, v_{i}\right)^{T} \mathbf{T}_{m,i} \mathbf{\Phi}_i \mathbf{b}\left(f_m, u_{1}, v_{1}\right)\right| \\ & =\frac{1}{N_{\mathrm{RIS}}}\left|\sum_{\mathrm{n}_1=0}^{N_1-1} \sum_{\mathrm{n}_2=0}^{N_2-1} e^{j \pi \xi_{m} \left[n_1 (\varsigma_i+\varsigma_1) +n_2 (\eta_i+\eta_1)\right]} \cdot e^{j (\phi_{i,n}- 2 \pi f_m \tau_{i,n})}\right|,
\end{split}
\end{equation}
where $n=N_2 n_1+n_2+1$, $\mathbf{T}_{m,i}=\operatorname{diag}(e^{-j 2 \pi f_m \boldsymbol{\tau}_i})$ is the reflection/transmission time delay matrix on the $m$-th subcarrier with $\boldsymbol{\tau}_i=\left[\tau_{i,1}, \tau_{i,2}, \ldots, \tau_{i,N_{\mathrm{RIS}}}\right]^T$.  To obtain maximum normalized array gain $\mathrm{g}_{\mathrm{fully},i}(f_m)$ over all frequencies, $\tau_{i,n}$ is designed~as
\begin{eqnarray}
\tau_{i,n}=\frac{1}{2 f_c}[n_1 (\varsigma_i+\varsigma_1) +n_2 (\eta_i+\eta_1)],
\end{eqnarray}
and $\phi_{i,n}$ can be set to arbitrary constants.

Consequently, the generated beams of the STAR-RIS with fully-connected structure will be perfectly steered to the target across the entire frequencies and thus the beam split is mitigated completely. Nonetheless, the cost and power consumption of TDs are much higher than that of PSs~\cite{ref27}. To further trade off the hardware cost and performance, we introduce a more cost-effective sub-connected structure, as presented in Fig. 3(b). To be specific, each side of the STAR-RIS element adopts double-layer PSs, and multiple STAR-RIS elements are combined into a sub-surface to share one TD. The first layer PS-r/PS-t is used to adjust the phase difference of the signals, and the second PS-r/PS-t is used to form beam. In this paper, the STAR-RIS is divided into $S = S_1 S_2$ sub-surfaces. Each sub-surface consists of $L = L_1 L_2$ elements, where $L_1 = N_1/S_1$, $L_2 = N_2/S_2$.  We denote $\tau_{i,s}$, $\boldsymbol{\Phi}_{i,1, s}$ and $\boldsymbol{\Phi}_{i,2, s}$, $\forall i \in \{ \mathbb{R}, \mathbb{T}\}$, as the time delay, first-layer and second-layer phase-shift at the $s$-th sub-surface of reflection/transmission side, respectively, where $s=s_1 S_2+s_2+1$. Then, the reflection/transmission array gain $\mathrm{g}_{\mathrm{sub},i}(f_m), \forall i \in \{ \mathbb{R}, \mathbb{T}\}$ of the STAR-RIS with sub-connected structure on the $m$-th subcarrier is computed by
\begin{eqnarray}
\begin{aligned}
&\mathrm{g}_{\mathrm{sub},i}\left(f_m\right)=\left|\sum_{s_1=0}^{S_1-1} \sum_{s_2=0}^{S_2-1} \right.\\ & \left.
\!\!\!\!\left\{\sum_{l_1=0}^{L_1-1} \sum_{l_2=0}^{L_2-1} \mathbf{h}_{m, i,s}\left[\boldsymbol{\Phi}_{i,2, s} e^{-j 2 \pi f_m \tau_{i,s}}\sum_{l_1=0}^{L_1-1} \sum_{l_2=0}^{L_2-1} \boldsymbol{\Phi}_{i,1, s} \mathbf{G}_{m, s}\right]\right\}\right|,
\end{aligned}
\end{eqnarray}
where $\mathbf{G}_{m, s}$ and $\mathbf{h}_{m,i, s}$ denote the channel from the BS to STAR-RIS and from the STAR-RIS to reflection/transmission user at $s$-th sub-surface on the $m$-th subcarrier, respectively.
Next, we optimize $\tau_{i,s}$, $\boldsymbol{\Phi}_{i,1, s}$ and $\boldsymbol{\Phi}_{i,2, s}$ to maximize the reflection/transmission array gain of the STAR-RIS on the $m$-th subcarrier. Due to $\sum_{l_1=0}^{L_1-1} \sum_{l_2=0}^{L_2-1} \boldsymbol{\Phi}_{i, 1, s} \mathbf{G}_{m, s}$ is a scalar,  maximizing $ \mathrm{g}_{\mathrm{sub},i}\left(f_m\right)$ can be divided into two subproblems, namely
\begin{eqnarray}
\max \sum_{l_1=0}^{L_1-1} \sum_{l_2=0}^{L_2-1} \boldsymbol{\Phi}_{i, 1, s} \mathbf{G}_{m, s},
\end{eqnarray}
and
\begin{eqnarray}
\max \sum_{l_1=0}^{L_1-1} \sum_{l_2=0}^{L_2-1} \mathbf{h}_{m, i,s} \boldsymbol{\Phi}_{i,2, s}.
\end{eqnarray}
The $(l_1, l_2 )$-th element of $\mathbf{G}_{m, s}$ is written as
\begin{eqnarray}
\begin{aligned}
& \mathbf{G}_{m, s,\left(l_1, l_2\right)}=\frac{\alpha_1}{\sqrt{N_{\mathrm{RIS}}}} e^{-j 2 \pi t_1 f_m} e^{j \pi \xi_m\left(n_1 \varsigma_1+n_2 \eta_1\right)} \\ & =\frac{\alpha_1}{\sqrt{N_{\mathrm{RIS}}}} e^{-j 2 \pi t_1 f_m} e^{j \pi \xi_m\left(\left(s_1 L_1+l_1\right) \varsigma_1+\left(s_2 L_2+l_2\right) \eta_1\right)}  \\ & =\frac{\alpha_1}{\sqrt{N_{\mathrm{RIS}}}} e^{-j 2 \pi t_1 f_m} e^{j \pi \xi_m\left(s_1 L_1 \varsigma_1+s_2 L_2 \eta_1\right)} e^{j \pi \xi_m\left(l_1 \varsigma_1+l_2 \eta_1\right)},
\end{aligned}
\end{eqnarray}
where $n_1=s_1L_1+l_1$, $n_2=s_2L_2+l_2$. To maximize (12), the $(l_1, l_2 )$-th element of $\boldsymbol{\Phi}_{i,1, s}$ should be designed as
\begin{eqnarray}
\boldsymbol{\Phi}_{i,1, s, (l_1, l_2)}=e^{-j \pi \left(l_1 \varsigma_1+l_2 \eta_1\right)}, \forall i \in \{ \mathbb{R}, \mathbb{T}\}.
\end{eqnarray}
Furthermore, $\sum_{l_1=0}^{L_1-1} \sum_{l_2=0}^{L_2-1} \boldsymbol{\Phi}_{i,1, s} \mathbf{G}_{m, s}$ is equal to
\begin{eqnarray}
\begin{aligned}
& \sum_{l_1=0}^{L_1-1} \sum_{l_2=0}^{L_2-1} \boldsymbol{\Phi}_{i,1, s} \mathbf{G}_{m, s}\\ & =\frac{\alpha_1 }{\sqrt{N_{\mathrm{RIS}}}} e^{-j 2 \pi t_1 f_m} e^{j \pi \xi_m\left(s_1 L_1 \varsigma_1+s_2 L_2 \eta_1\right)} e^{-j \pi\left(\xi_m-1\right) \frac{\left(L_1-1\right) \varsigma_1}{2}}\\
&e^{-j \pi\left(\xi_m-1\right) \frac{\left(L_2-1\right) \eta_1}{2}} \Xi_{L_1}\left(\left(\xi_m-1\right) \varsigma_1\right) \Xi_{L_2}\left(\left(\xi_m-1\right) \eta_1\right).
\end{aligned}
\end{eqnarray}
Similarly, the $(l_1, l_2 )$-th element of $\mathbf{h}_{m,i, s}$ can be written~as
\begin{equation}
\begin{split}
& \mathbf{h}_{m, i,s,\left(l_1, l_2\right)}=\frac{\alpha_i}{\sqrt{N_{\mathrm{RIS}}}} e^{-j 2 \pi t_i f_m} e^{j \pi \xi_m\left(s_1 L_1 \varsigma_i+s_2 L_2 \eta_i\right)} e^{j \pi \xi_m\left(l_1 \varsigma_i+l_2 \eta_i\right)}.
\end{split}
\end{equation}
And thus the $(l_1, l_2 )$-th element of $\boldsymbol{\Phi}_{i,2, s}$ should be
\begin{eqnarray}
\boldsymbol{\Phi}_{i,2, s, (l_1, l_2)}=e^{-j \pi \left(l_1 \varsigma_i+l_2 \eta_i\right)}, \forall i \in \{ \mathbb{R}, \mathbb{T}\}.
\end{eqnarray}
Then, we can obtain
\begin{eqnarray}
\begin{aligned}
& \sum_{l_1=0}^{L_1-1} \sum_{l_2=0}^{L_2-1} \mathbf{h}_{m,i, s} \boldsymbol{\Phi}_{i,2, s} =\\
&\frac{\alpha_i}{\sqrt{N_{\mathrm{RIS}}}} e^{-j 2 \pi t_i f_m} e^{j \pi \xi_m\left(s_1 L_1 \varsigma_i+s_2 L_2 \eta_i\right)} e^{-j \pi\left(\xi_m-1\right) \frac{\left(L_1-1\right) \varsigma_i}{2}}\\
&e^{-j \pi\left(\xi_m-1\right) \frac{\left(L_2-1\right) \eta_i}{2}} \Xi_{L_1}\left(\left(\xi_m-1\right) \varsigma_i\right) \Xi_{L_2}\left(\left(\xi_m-1\right) \eta_i\right).
\end{aligned}
\end{eqnarray}
By ignoring the modulus constraint, the reflection/transmission array gain $\mathrm{g}_{\mathrm{sub},i}(f_m), \forall i \in \{ \mathbb{R}, \mathbb{T}\}$ can be derived as
\begin{eqnarray}
\begin{aligned}
& \mathrm{g}_{\mathrm{sub},i}\left(f_m\right)= \frac{1}{N_{\mathrm{RIS}}} \Xi_{L_1}\left(\left(\xi_m-1\right) \varsigma_1\right) \Xi_{L_2}\left(\left(\xi_m-1\right) \eta_1\right)\\& \Xi_{L_1}\left(\left(\xi_m-1\right) \varsigma_i\right) \Xi_{L_2}\left(\left(\xi_m-1\right) \eta_i\right)\left|\sum_{s_1=0}^{S_1-1} \sum_{s_2=0}^{S_2-1} \right.\\ & \left.
\!\!\!\left\{e^{-j 2 \pi f_m \tau_{i,s}}\cdot e^{j \pi \xi_m\left[\left(s_1 L_1-\frac{\left(L_1-1\right)} {2}\right)(\varsigma_i+\varsigma_1)+\left(s_2 L_2-\frac{\left(L_2-1\right)} {2}\right)(\eta_i+\eta_1) \right]}\right\}\right|.
\end{aligned}
\end{eqnarray}
To relieve the beam split effect, the  reflection/transmission time delay $\tau_{i,s}$ of the $s$-th sub-surface should satisfy
\begin{eqnarray}
\begin{aligned}
\tau_{i,s}
&=\frac{1}{2 f_c}\left[\left(s_1 L_1-\frac{\left(L_1-1\right)} {2}\right)(\varsigma_i+\varsigma_1)+\right.\\ & \left.
\left(s_2 L_2-\frac{\left(L_2-1\right)} {2}\right)(\eta_i+\eta_1)\right], \forall i \in \{ \mathbb{R}, \mathbb{T}\}.
\end{aligned}
\end{eqnarray}
Finally, based on the obtained $\tau_{i,s}$, $\boldsymbol{\Phi}_{i,1, s}$ and $\boldsymbol{\Phi}_{i,2, s}$, we have
\begin{eqnarray}
\begin{aligned}
& \mathrm{g}_{\mathrm{sub},i}\left(f_m\right)= \frac{S_1 S_2}{N_{\mathrm{RIS}}} \Xi_{L_1}\left(\left(\xi_m-1\right) \varsigma_1\right) \Xi_{L_2}\left(\left(\xi_m-1\right) \eta_1\right)\\& \Xi_{L_1}\left(\left(\xi_m-1\right) \varsigma_i\right) \Xi_{L_2}\left(\left(\xi_m-1\right) \eta_i\right),\forall i \in \{ \mathbb{R}, \mathbb{T}\}.
\end{aligned}
\end{eqnarray}

Furthermore, Fig. 4 evaluates the normalized array gain with different structures, where $f_{c}=100$ GHz, $B=10$GHz, $M=128$, $N_{\rm RIS}=16 \times 16$ and $S=4 \times 4$. We can find that the conventional STAR-RIS structure without TD is unable to produce the desired phase-shifts at different frequencies, which causes sever array gain loss. Besides, it is observed that the STAR-RIS with sub-connected structure is more energy efficient due to the drastic reduction in the number of TDs, and a sub-optimal solution can be obtained.
\begin{figure}[htbp]
\centering
    \label{RIS_subcarrier} 
    \includegraphics[width=8.5cm,height=5.5cm]{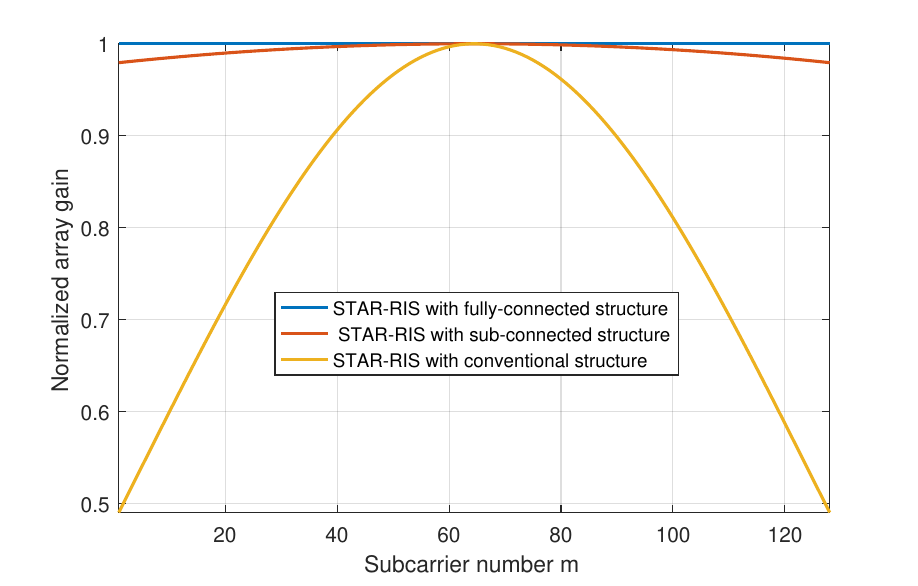}
	\caption{Normalized array gain under different structures.}
\end{figure}

\section{System Model and Problem Formulation}
In this section, we study a multiple STAR-RISs-assisted THz communication system, where the TD-based hybrid analog/digital structure at the BS and the TD-based sub-connected structure at each SATR-RIS are considered, respectively. Based on the proposed scheme, a sum rate maximization problem is  formulated.
\subsection{System Model}
We investigate a multiple STAR-RISs-assisted THz communication system as illustrate in Fig. 5. The BS deploys $N_t$ antennas and $N_{{\rm{RF}}}$ $(N_t \geq N_{{\rm{RF}}})$ RF chains to serve $K$ single-antenna users assisted by $R$ STAR-RISs. Let $\mathcal{R}=\{1, \cdots, R\}$ denote the set of STAR-RISs. We assume that all STAR-RISs have the same size, i.e., $N_{\rm RIS}$. And we define $\mathcal{K}=\{1,2,\cdots,K\}$ as the set of users, wherein $\mathcal{K}_{\mathbb{R}}=\{1,2,\cdots,K_0\}$ and $\mathcal{K}_{\mathbb{T}}=\{K_0+1,\cdots,K\}$ as the set of users located on the reflection and transmission side, respectively. To simply the system management, we assume that each STAR-RIS serves one reflection user and one transmission user each time according to its coverage range, and thus we have $K=2R$. Meanwhile, the sub-connected structure at each STAR-RIS is applied, and each STAR-RIS is divided into $S$ sub-surfaces and each sub-surface consists of $L$ elements.

The channel matrix $\mathbf{G}_{r, m} \in \mathbb{C}^{N_{\rm{RIS}} \times N_{t}}$ between the BS and the $r$-th STAR-RIS can be expressed as
\begin{eqnarray}
\mathbf{G}_{r,m}= \alpha_{b,r} e^{-j 2 \pi t_{b,r} f_{m}} \mathbf{b}\left(f_m, u_{b,r}, v_{b,r}\right) \mathbf{a}\left(f_m, \theta_{b,r}\right)^{H},
\end{eqnarray}
where
\begin{equation}
\begin{split}
\mathbf{a}\left(f_m, \theta_{b,r}\right)
\!=\!\frac{1}{\sqrt{N_{t}}}\left[1, \!\ldots,\! e^{j \pi \xi_{m} n_{t} \sin \theta_{b,r}},\! \ldots,\! e^{\left.j  \pi \xi_{m}\left(N_{t}\!-\!1\right) \sin \theta_{b,r}\right)}\right]^{T},
\end{split}
\end{equation}
is the steering vector of the BS, $t_{b,r}$ and $\alpha_{b,r}$ are the path delay and gain from the BS to the $r$-th STAR-RIS, respectively.
$\theta_{b,r} \in[-\pi / 2, \pi / 2]$ is the AoD of the BS, $u_{b,r}\in[-\pi / 2, \pi / 2]$ and $v_{b,r} \in[0, \pi]$ are the azimuth and elevation AOA of the $r$-th STAR-RIS, respectively.
\begin{figure}[htbp]
\centering
    \label{RIS_subcarrier} 
    \includegraphics[width=8cm,height=5cm]{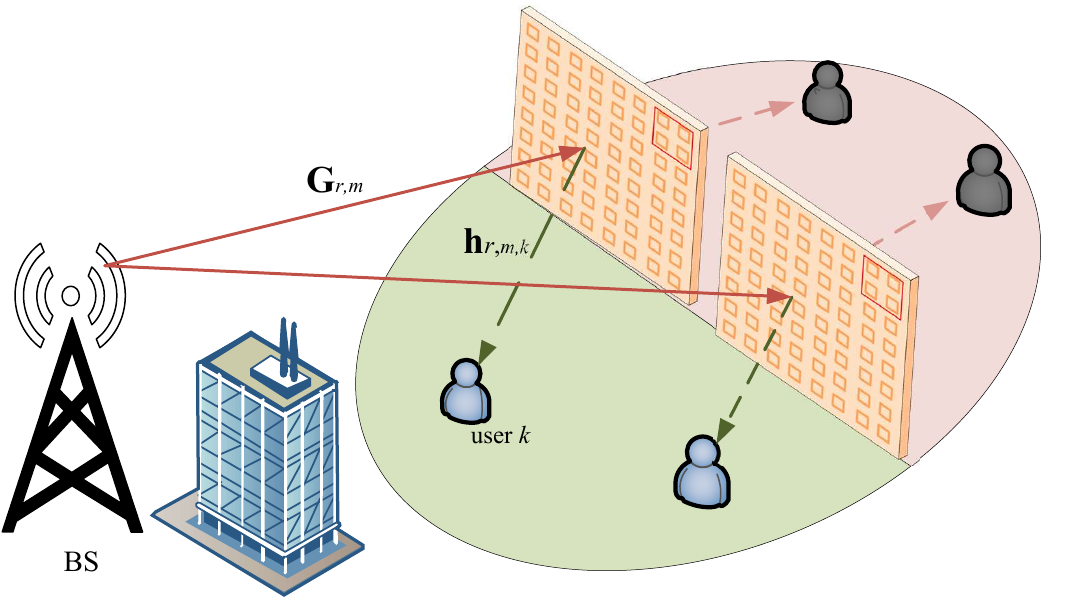}
	
	\caption{The multiple STAR-RISs-assisted THz system.}
\end{figure}

Similarly, the frequency-domain channel vector $\mathbf{h}_{r, m, k} \in \mathbb{C}^{1 \times N_{\rm{RIS}}}$ between the $r$-th STAR-RIS and the $k$-th user, $ \forall k \in \mathcal{K}$, is modeled as
\begin{eqnarray}
\mathbf{h}_{r, m, k}=\alpha_{r, k} e^{-j 2 \pi t_{r,k} f_{m}} \mathbf{b}\left(f_m, u_{r,k}, v_{r,k}\right)^{T},
\end{eqnarray}
where $t_{r,k}$ and $\alpha_{r, k}$ represent the path delay and gain from the $r$-th STAR-RIS to the $k$-th user, respectively.  $u_{r,k}\in[-\pi / 2, \pi / 2]$ and $v_{r,k} \in[0, \pi]$ denote the azimuth and elevation angle of AOD at the $r$-th STAR-RIS, respectively.

Therefore, the equivalent channel $\boldsymbol{\hbar}_{m, k}$ between the BS and user $k$, $ \forall k \in \mathcal{K}$, on the $m$-th subcarrier is given by
\begin{eqnarray}
\begin{aligned}
\boldsymbol{\hbar}_{m, k}&=\sum_{r=1}^{R} \mathbf{h}_{r, m, k} \mathbf{A}_{i,r} \boldsymbol{\Phi}_{i,r,2} \mathbf{T}_{i,r,m} \boldsymbol{\Phi}_{i,r,1} \mathbf{G}_{r, m}\\&
\overset{\mathrm{(a)}}{=} \mathbf{h}_{ m, k} \mathbf{A}_{i} \boldsymbol{\Phi}_{i,2} \mathbf{T}_{i,m} \boldsymbol{\Phi}_{i,1} \mathbf{G}_{m}\\&
\overset{\mathrm{(b)}}{=} \mathbf{h}_{ m, k} \mathbf{A}_{i} \boldsymbol{\Phi}_{i,m} \mathbf{G}_{m}, \forall i \in \{ \mathbb{R}, \mathbb{T}\},
 \end{aligned}
\end{eqnarray}
where $\mathbf{A}_{i,r}=\operatorname{diag}\left(\beta_{i,r,1}, \cdots, \beta_{i,r,N_{\rm{RIS}}}\right), \forall r \in \mathcal{R}$ is the diagonal amplitude matrix of the $r$-th STAR-RIS, $\boldsymbol{\Phi}_{i,r,1}=\operatorname{diag}\left(e^{j \phi_{i,r,1,1}}, \cdots, e^{j \phi_{i,r,1,n}}, \cdots, e^{j \phi_{i,r,1,N_{\rm{RIS}}}}\right)$, $\boldsymbol{\Phi}_{i,r,2}=\operatorname{diag}\left(e^{j \phi_{i,r,2,1}}, \cdots, e^{j \phi_{i,r,2,n}}, \cdots, e^{j \phi_{i,r,2,N_{\rm{RIS}}}}\right)$ and $\mathbf{T}_{i,r,m}$ are the first-layer, second-layer diagonal reflection/transmission coefficients matrix and time delay matrix of the STAR-RIS, respectively. (a) holds by setting $\mathbf{h}_{m,k}=\left[\mathbf{h}_{1,m,k}^{T}, \ldots, \mathbf{h}_{r, m, k}^{T}, \ldots, \mathbf{h}_{R, m, k}^{T}\right]^{T}$, $\mathbf{G}_{m}=\left[ \mathbf{G}_{1,m}^{T}, \ldots, \mathbf{G}_{r, m}^{T}, \ldots, \mathbf{G}_{R, m}^{T}\right]^{T}$, $\mathbf{A}_{i}=\operatorname{diag}(\mathbf{A}_{i,1}, \ldots, \mathbf{A}_{i, r}, \ldots, \mathbf{A}_{i, R})$, $\boldsymbol{\Phi}_{i,1}=\operatorname{diag}(\boldsymbol{\Phi}_{i,1,1}, \ldots, \boldsymbol{\Phi}_{i,r,1}, \ldots, \boldsymbol{\Phi}_{i,R,1})$, $\boldsymbol{\Phi}_{i,2}=\operatorname{diag}(\boldsymbol{\Phi}_{i,1,2}, \ldots, \boldsymbol{\Phi}_{i,r,2}, \ldots, \boldsymbol{\Phi}_{i,R,2})$ and $\mathbf{T}_{i,m}=\operatorname{diag}(\mathbf{T}_{i,1,m}, \ldots, \mathbf{T}_{i,r,m}, \ldots, \mathbf{T}_{i,R,m})$, (b) holds by setting $\boldsymbol{\Phi}_{i,m}=\boldsymbol{\Phi}_{i,2} \mathbf{T}_{i,m} \boldsymbol{\Phi}_{i,1}$.
In particular, for independent  phase-shift STARS-RIS, the energy conservation constraint is required to be satisfied, namely
\begin{eqnarray}
\sum_{i \in \{ \mathbb{R}, \mathbb{T}\}}\beta_{i,r,n}^{2}\leq 1, \forall r \in \mathcal{R}, n=1,2, \ldots, N_{\rm RIS}.
\end{eqnarray}
For the BS, each RF chain is connected to $K_{\rm{T}}$ TDs and each TD is connected to $P = N_{t}/ K_{\rm{T}}$ PSs. Thus the received signal of the $k$-th user, $ \forall k \in \mathcal{K}$, on the $m$-th subcarrier is given by
\begin{eqnarray}\label{ZP}
y_{m, k}=
\boldsymbol{\hbar}_{m, k} \sum_{j=1}^{K} \mathbf{F}_{\rm{A}} \mathbf{F}_{m}^{\rm{td}}\mathbf{d}_{m, j} s_{m, j}+n_{m, k},
\end{eqnarray}
where $\mathbf{F}_{\rm{A}}\in \mathbb{C}^{N_{t} \times K_{\rm{T}}N_{\rm{RF}}}=[\mathbf{F}_{A,1}, \cdots, \mathbf{F}_{A,l}, \cdots,\mathbf{F}_{A,N_{\rm{RF}}}]$ is analog beamforming realized by PSs,  $\mathbf{F}_{A,l}\in \mathbb{C}^{N_{t} \times K_{\rm{T}}}= \operatorname{diag}([\mathbf{f}_{A,l,1},\mathbf{f}_{A,l,2},\cdots,\mathbf{f}_{A,l,K_{\rm{T}}}])$ denotes the analog beamforming matrix achieved by the PSs connecting to the $l$-th RF chain. $\mathbf{F}_{m}^{\rm{td}}\in \mathbb{C}^{ K_{\rm{T}}N_{\rm{RF}}\times N_{\rm{RF}}}=\operatorname{diag}\left(\left[e^{-j 2 \pi f_{m} \mathbf{z}_{1}}, e^{-j 2 \pi f_{m} \mathbf{z}_{2}},\cdots, e^{-j 2 \pi f_{m} \mathbf{z}_{N_{\rm{RF}}}}\right]\right)$ is the time delay matrix, where $\mathbf{z}_{l} \in \mathbb{C}^{K_{\rm{T}} \times 1}=\left[z_{l, 1}, z_{l, 2}, \cdots, z_{l, K_{\rm{T}}}\right]^{ T}$ denotes time delay vector generated by $K_{\rm{T}}$ TDs via the $l$-th RF chain.
$\mathbf{d}_{m, k}\in \mathbb{C}^{N_{\rm{RF}} \times 1}$ is the digital beamforming vector. In addition, $n_{m,k} \sim \mathcal{C} \mathcal{N}\left(0, \sigma_{m,k}^{2}\right)$ denotes the additive white Gaussian noise (AWGN) at the $k$-th user on the $m$-th subcarrier, and the transmit symbol $s_{m, k}$ satisfies $\mathbb{E}\left[\left|s_{m,k}\right|^{2}\right]=1$.
Then, the $k$-th user's SINR, $ \forall k \in \mathcal{K}$, on the $m$-th subcarrier is computed by
\begin{eqnarray}
\gamma_{m,k}=\frac{\left|\boldsymbol{\hbar}_{m, k} \mathbf{F}_{\rm{A}} \mathbf{F}_{m}^{\rm{td}}\mathbf{d}_{m, k}\right|^{2}}{\sum_{j=1, j \neq k}^{K}\left|\boldsymbol{\hbar}_{m, k} \mathbf{F}_{\rm{A}} \mathbf{F}_{m}^{\rm{td}}\mathbf{d}_{m, j}\right|^{2}+\sigma_{m, k}^{2}},
\end{eqnarray}
and the corresponding sum rate is
\begin{eqnarray}
R_{\rm{sum}}=\sum_{k=1}^{K} \sum_{m=1}^{M} \log _{2}\left(1+\gamma_{m, k}\right).
\end{eqnarray}
\subsection{Problem Formulation}
We consider to maximize the sum rate via jointly optimizing the analog beamforming $\mathbf{F}_{\rm{A}}$, digital beamforming $\mathbf{d}_{m, k}$, time delay $\mathbf{F}_{m}^{\rm{td}}$  at the BS as well as double-layer phase-shift coefficients $\boldsymbol{\Phi}_{i,1}$, $\boldsymbol{\Phi}_{i,2}$, time delays $\mathbf{T}_{i,m}$ and amplitude coefficients $\mathbf{A}_{i}$, $\forall i \in \{ \mathbb{R}, \mathbb{T}\}$, at the STAR-RISs.  Therefore, the optimization problem is formulated as
\begin{subequations}\label{OptA}
\begin{align}
\mathrm{P0}:&\max _{\boldsymbol{\Phi}_{i,m}, \mathbf{A}_{i},\mathbf{F}_{\rm{A}},\mathbf{F}_{m}^{\rm{td}}, \mathbf{d}_{m, k}} R_{\rm{sum}}\label{OptA0}\\
{\rm{s.t.}}\;\;&\sum_{\mathrm{k}=1}^{K} \sum_{m=1}^{M}\left\|\mathbf{F}_{\rm{A}}\mathbf{F}_{m}^{\rm{td}} \mathbf{d}_{m, k}\right\|^{2} \leq P_{\text {max }} \label{OptA1},\\
&\sum_{i \in \{ \mathbb{R}, \mathbb{T}\}}\beta_{i,r,n}^{2}\leq 1, \forall r \in \mathcal{R}, n=1,2, \ldots, N_{\rm RIS}\label{OptA2},\\
&\left|\mathbf{f}_{A,l, k_{t}}\right|=\frac{1}{\sqrt{N_{t}}}, l=1,2, \ldots N_{\rm{RF}}, k_{t}=1,2, \ldots, K_{\rm{T}},
\end{align}
\end{subequations}
where $P_{\text {max }}$ is the maximum transmit power. (31b) is the power constraint, (31c) is the energy conservation constraint for independent phase-shift STAR-RIS, and (31d) is imposed by the hardware limitation of PSs. Due to the non-convex objective function (31a) and the constant modulus constraint (31d), the optimization problem is challenging to be tackled directly. Next, an effective algorithm is proposed to address it.
\section{Solution of the Optimization Problem}
In this section, the analog beamforming and time delays of the BS are first designed based on different STAR-RISs' physical directions, and then we derive time delays, the double-layer phase-shifts of each STAR-RIS according to the analysis in Section II-B. Finally, we propose an alternative optimization method to calculate the digital beamforming of the BS and reflection/transmission amplitude of each STAR-RIS.
\subsection{Solution of $\mathbf{F}_{\rm{A}}$, $\mathbf{F}_{m}^{td}$, $\boldsymbol{\Phi}_{i,m}$ }
We define the frequency-dependent analog beamforming matrix as $\mathbf{F}_m=\mathbf{F}_{\rm{A}}\mathbf{F}_{m}^{\rm{td}}$, and thus the $l$-th column of the analog beamforming vector is represent as $\mathbf{\bar{f}}_{l,m}=\mathbf{F}_{A,l}e^{-j 2 \pi f_{m} \mathbf{z}_{l}}$, which is used to serve the $r$-th STAR-RIS on the $m$-th subcarrier.
Based on \cite{ref27_0}, \cite{ref27_1}, PSs are applied to generate beams towarding to STAR-RISs' physical directions, and TDs are introduced to rotate deviated beams at different subcarriers. Therefore, $\mathbf{F}_{A, l}$ can be expressed as
\begin{eqnarray}\label{ZN}
\mathbf{F}_{A,l}=\operatorname{diag}\left(\left[\mathbf{a}_{1 \rightarrow P}\left(\varepsilon_{b,r}^{c}\right), \ldots, \mathbf{a}_{\left(K_{\rm T}-1\right) P \rightarrow K_{\rm T} P}\left(\varepsilon_{b,r}^{c}\right)\right]\right),
\end{eqnarray}
 where $\varepsilon_{b,r}^{c}= \xi_{c} \sin \theta_{b,r}$.
However, the beam generated on the $m$-th subcarrier will steer to the frequency-dependent direction $\varepsilon_{b,r}^{m}$, namely
\begin{eqnarray}
\varepsilon_{b,r}^{m}=(f_{c}/f_{m}) \varepsilon_{b,r}^{c}.
\end{eqnarray}
To relieve the beam split effect, TDs are applied to provide an extra phase-shift $\varrho_{l,m}$, where $\varrho_{l,m}=(\xi_m -1)P\varepsilon_{b,r}^{c}$.
Therefore, for the $l$-th beam, the time delays is
\begin{eqnarray}\label{ZO}
e^{-j 2 \pi f_{m} \mathbf{z}_{l}}=\left[1, e^{j \pi \varrho_{l,m}}, e^{j \pi 2 \varrho_{l,m}}, \cdots, e^{j \pi(K_{\rm{T}}-1) \varrho_{l,m}}\right]^{T}.
\end{eqnarray}
The time delays vector $\mathbf{z}_{l}$ is calculated as
\begin{eqnarray}
\mathbf{z}_{l} \in \mathbb{C}^{K_{\rm{T}} \times 1}=\left[0, b_{l} T_{c}, \ldots, b_{l} T_{c}\left(K_{\rm{T}}-1\right)\right]^{T},
\end{eqnarray}
where $b_{l}=-\frac{P \sin \varepsilon_{b,r}^{c}}{2}$ is the number of periods that need to delay for the $l$-th path component, $T_{c}$ is the corresponding period of the central~frequency.

Next, we design $\boldsymbol{\Phi}_{i,m}=\boldsymbol{\Phi}_{i,2} \mathbf{T}_{i,m} \boldsymbol{\Phi}_{i,1}, \forall i \in \{ \mathbb{R}, \mathbb{T}\}$, including the time delay matrix $\mathbf{T}_{i,m}$, and double-layer phase-shift coefficients matrices $\boldsymbol{\Phi}_{i, 1}$, $\boldsymbol{\Phi}_{i, 2}$. As analyzed in Section II-B, the time delay and phase-shift are just related to the user' location.
Thus, we can extend it to the multi-user scenario by deploying multiple STAR-RISs. According to (15), (18) and (21), we can obtain the first-layer phase-shift $\boldsymbol{\Phi}_{i,r,1}$, second-layer phase-shift $\boldsymbol{\Phi}_{i,r,2}$ and time delay $\mathbf{T}_{i,r,m}$ of the $r$-th STAR-RIS, respectively. The $(l_1, l_2)$-th element within the $s$-th sub-surface of $\boldsymbol{\Phi}_{i,r,1}$ should be designed as
\begin{eqnarray}
\boldsymbol{\Phi}_{i,r,1, s, (l_1, l_2)}=e^{-j \pi \left(l_1 \varsigma_{b,r}+l_2 \eta_{b,r}\right)},
\end{eqnarray}
where $\varsigma_{b,r}=\sin u_{b,r} \sin v_{b,r}$, $\eta_{b,r}=\cos u_{b,r}$. Similarly, the $(l_1, l_2)$-th element within the $s$-th sub-surface of $\boldsymbol{\Phi}_{i,r,2}$ should be given~by
\begin{eqnarray}
\boldsymbol{\Phi}_{i,r,2, s, (l_1, l_2)}=e^{-j \pi \left(l_1 \varsigma_{r,k}+l_2 \eta_{r,k}\right)},
\end{eqnarray}
where $\varsigma_{r,k}=\sin u_{r,k} \sin v_{r,k}$, $\eta_{r,k}=\cos u_{r,k}$, $ \forall k \in \mathcal{K}$. Then, we denote $\tau_{i,r,s}$ as the reflection/transmission time delay on the $s$-th sub-surface of $r$-th STAR-RIS, which should satisfy
\begin{equation}
\begin{split}
\tau_{i,r,s}=\frac{1}{2 f_c}\left[\left(s_1 L_1-\frac{\left(L_1-1\right)} {2}\right)\varsigma_{r}+\left(s_2 L_2-\frac{\left(L_2-1\right)} {2}\right)\eta_{r}\right],
\end{split}
\end{equation}
where $\varsigma_{r}=\varsigma_{b,r}+\varsigma_{r,k}$, $\eta_{r}=\eta_{b,r}+\eta_{r,k}$. We further obtain the time delay matrix on the $s$-th sub-surface of the $r$-th STAR-RIS on the $m$-th subcarrier, i.e.,
\begin{eqnarray}
\mathbf{T}_{i,r,s,m}=I_L \otimes \bar{\tau}_{i,r,s,m}, \;\; \bar{\tau}_{i,r,s,m}=e^{-j 2 \pi f_m \tau_{i,r,s}}.
\end{eqnarray}
Consequently, the time delay matrix of the $r$-th STAR-RIS on the $m$-th subcarrier is stated as $\mathbf{T}_{i,r,m}=\operatorname{diag}\left(\mathbf{T}_{i,r,1,m}, \cdots, \mathbf{T}_{i,r,s,m}, \cdots, \mathbf{T}_{i,r,S,m}\right)$.

After obtaining $\mathbf{F}_{\rm{A}}$, $\mathbf{F}_{m}^{\rm{td}}$, $\boldsymbol{\Phi}_{i,m}$, the function $R_{\rm{sum}}$ still has the expression of sum-of logarithmic function for $\gamma_{m,k}$, making the optimization problem P0 difficult to solve. Next, we adopt the fractional programming (FP) algorithm to tackle the non-convex objective function. Specifically, we employ the LDR to transform the objective function as~\cite{ref28},~\cite{ref29}
\begin{eqnarray}
\begin{split}
f(\mathbf{A}_{i}, \mathbf{d}_{m,k}, \boldsymbol{\rho})=\sum_{k=1}^{K} \sum_{m=1}^{M} \ln \left(1+\rho_{m, k}\right)-\sum_{k=1}^{K} \sum_{m=1}^{M} \rho_{m, k}+\\
\sum_{k=1}^{K} \sum_{m=1}^{M}\left(1+\rho_{m, k}\right) f_{m, k}(\mathbf{A}_{i}, \mathbf{d}_{m,k}),
\end{split}
\end{eqnarray}
where $\boldsymbol{\rho}=\left[\mathbf{\rho}_{1,1}, \mathbf{\rho}_{1,2}, \ldots, \mathbf{\rho}_{1, K}, \mathbf{\rho}_{2,1}, \mathbf{\rho}_{2,2}, \ldots, \mathbf{\rho}_{M, K}\right]^{T}$ is an auxiliary variable, and $f_{m, k}(\mathbf{A}_{i}, \mathbf{d}_{m,k})$ is denoted as
\begin{eqnarray}
f_{m, k}(\mathbf{A}_{i}, \mathbf{d}_{m,k})=\frac{\left|\widehat{\mathbf{h}}_{m, k} \mathbf{d}_{m, k}\right|^{2}}{\sum_{j=1}^{K}\left|\widehat{\mathbf{h}}_{m, k} \mathbf{d}_{m, j}\right|^{2}+\sigma_{m, k}^{2}},
\end{eqnarray}
where $\widehat{\mathbf{h}}_{m, k}=\boldsymbol{\hbar}_{m, k} \mathbf{F}_{m} $ is the equivalent channel.

Therefore, the problem $\mathrm{P0}$  is equivalent to the following problem:
\begin{subequations}\label{OptA}
\begin{align}
\mathrm{P1}:&\max _{\mathbf{A}_{i}, \mathbf{d}_{m, k}, \boldsymbol{\rho}} f(\mathbf{A}_{i}, \mathbf{d}_{m,k}, \boldsymbol{\rho})\label{OptA0}\\
{\rm{s.t.}}\;\;&\sum_{\mathrm{k}=1}^{K} \sum_{m=1}^{M}\left\|\mathbf{F}_m \mathbf{d}_{m, k}\right\|^{2} \leq P_{\text {max }} \label{OptA1},\\
&\sum_{i \in \{ \mathbb{R}, \mathbb{T}\}}\beta_{i,r,n}^{2}\leq 1, \forall r \in \mathcal{R}, n=1,2, \ldots, N_{\rm RIS}\label{OptA2}.
\end{align}
\end{subequations}
Then, we utilize a joint alternative optimization scheme to optimize $\boldsymbol{\rho}$, $\mathbf{d}_{m,k}$ and $\mathbf{A}_{i}$ iteratively. By introducing another two auxiliary variables $\boldsymbol{\varpi}$ and $\boldsymbol{\epsilon}$, variables $\boldsymbol{\rho}$, $\boldsymbol{\varpi}$, $\mathbf{d}_{m,k}$, $\boldsymbol{\epsilon}$ and $\mathbf{A}_{i}$ are alternately updated until convergence. The details are summarized as the following subsections.
\subsection{Solution of $\mathbf{\rho}_{m, k}$ }
For given digital beamforming $\mathbf{d}_{m, k}$ and the reflection/transmission amplitude
coefficients $\mathbf{A}_{i}$, the objective function $f(\mathbf{A}_{i}, \mathbf{d}_{m,k}, \boldsymbol{\rho})$ is a convex function with respect to the variables $\mathbf{\rho}_{m, k}$, $\forall m \in \left[1,2, \cdots, M\right]$, $ \forall k \in \mathcal{K}$. Thus, the optimal solution of $\boldsymbol{\rho}$ is calculated by setting $\partial f / \partial \rho_{m, k}=0$, namely
\begin{eqnarray}
\rho_{m, k}^{*}=\frac{\left|\widehat{\mathbf{h}}_{m, k} \mathbf{d}_{m, k}\right|^{2}}{\sum_{j=1, j \neq k}^{K}\left|\widehat{\mathbf{h}}_{m, k} \mathbf{d}_{m, j}\right|^{2}+\sigma_{m, k}^{2}}.
\end{eqnarray}
\subsection{Solution of $\mathbf{d}_{m, k}$ }
For given $\boldsymbol{\rho}$ and $\mathbf{A}_{i}$, P1 is simplified as
\begin{subequations}\label{OptA}
\begin{align}
\;\;\mathrm{P2}:&\max _{\mathbf{d}_{m,k}}\sum_{k=1}^{K} \sum_{m=1}^{M}\left(1+\rho_{m,k}\right) f_{m, k}(\mathbf{A}_{i}, \mathbf{d}_{m,k})\label{OptA0}\\
{\rm{s.t.}}\;\;&\sum_{\mathrm{k}=1}^{K} \sum_{m=1}^{M}\left\|\mathbf{F}_{m} \mathbf{d}_{m, k}\right\|^{2} \leq P_{\text {max }}\label{OptA1}.
\end{align}
\end{subequations}
Unfortunately, since the objection function in (44a) is a sum of multiple  fractions, $\mathrm{P2}$ is still difficult to solve directly. Next, the objective function can be further equivalently transformed via MCQT~\cite{ref29}, namely
\begin{equation}
\begin{split}
g_1(\mathbf{d}_{m,k}, \boldsymbol{\varpi})&=\sum_{\mathrm{k}=1}^K \sum_{\mathrm{m}=1}^{\mathrm{M}} 2 \sqrt{\left(1+\rho_{m, k}\right)} \operatorname{Re}\left\{\varpi_{m, k}^{H} \widehat{\mathbf{h}}_{m, k} \mathbf{d}_{m, k}\right\}\\
&-\sum_{\mathrm{k}=1}^K \sum_{\mathrm{m}=1}^{\mathrm{M}} \varpi_{m, k}^{H}\left(\sum_{\mathrm{j}=1}^K\left|\widehat{\mathbf{h}}_{m, k} \mathbf{d}_{m, j}\right|^2+\sigma_{m, k}^2\right) \varpi_{m, k},
\end{split}
\end{equation}
where  $\boldsymbol{\varpi}=\left[\varpi_{1,1}, \varpi_{1,2}, \ldots, \varpi_{1, K}, \varpi_{2,1}, \varpi_{2,2}, \ldots, \varpi_{M, K}\right]$ is the auxiliary variable. Similarly, given other variables, $g_1$ is a convex function with respect to $\varpi_{m, k}$, $\forall m \in \left[1,2, \cdots, M\right]$, $ \forall k \in \mathcal{K}$. The optimal $\boldsymbol{\varpi}$ is obtained by setting $\partial g_1 / \partial \varpi_{m, k}=0$, namely
\begin{eqnarray}
\varpi_{m, k}^{*}=\frac{\sqrt{\left(1+\rho_{m, k}\right)}\left|\widehat{\mathbf{h}}_{m, k} \mathbf{d}_{m, k}\right|}{\sum_{j=1}^{K}\left|\widehat{\mathbf{h}}_{m, k} \mathbf{d}_{m, j}\right|^{2}+\sigma_{m, k}^{2}}.
\end{eqnarray}
Given $\boldsymbol{\rho}$, $\boldsymbol{\varpi}$
and $\mathbf{A}_{i}$,  the optimization problem with respect to $\mathbf{d}_{m,k}$ is simplified as
\begin{subequations}\label{OptA}
\begin{align}
\;\;\mathrm{P3}:&\max _{\mathbf{d}_{m,k}}g_1(\mathbf{d}_{m,k})\label{OptA0}\\
{\rm{s.t.}}\;\;&\sum_{\mathrm{k}=1}^{K} \sum_{m=1}^{M}\left\|\mathbf{F}_m \mathbf{d}_{m, k}\right\|^{2} \leq P_{\text {max }}\label{OptA1}.
\end{align}
\end{subequations}
Next, let
$\mathbf{d}=\left[\mathbf{d}_{1,1}^{T}, \mathbf{d}_{1,2}^{T}, \ldots, \mathbf{d}_{1, K}^{T}, \mathbf{d}_{2,1}^{T}, \mathbf{d}_{2,2}^{T}, \ldots, \mathbf{d}_{M, K}^{T}\right]^{T}$ and we define
\begin{eqnarray}
\begin{aligned}
\!\!\!\mathbf{e}_m \in \mathbb{C}^{N_{\mathrm{RF}} \times N_{\mathrm{RF}}}=\sum_{k=1}^{K} \widehat{\mathbf{h}}_{m, k}^{H} \varpi_{m, k}^{H} \varpi_{m, k} \widehat{\mathbf{h}}_{m, k},  \mathbf{E}_m=I_K \otimes \mathbf{e}_m,
\end{aligned}
\end{eqnarray}
\begin{eqnarray}
\begin{aligned}
\boldsymbol{v}_{m, k} \in \mathbb{C}^{N_{\mathrm{RF}} \times 1}=\sqrt{\left(1+\rho_{m, k}\right)} \widehat{\mathbf{h}}_{m, k}^{H} \varpi_{m, k},
\end{aligned}
\end{eqnarray}
\begin{eqnarray}
\begin{aligned}
Y=\sum_{\mathrm{k}=1}^K \sum_{\mathrm{m}=1}^{\mathrm{M}} \varpi_{m, k}^{H} \sigma_{m, k}^2 \varpi_{m, k}.
\end{aligned}
\end{eqnarray}
Then,  by substituting $(48)-(50)$ into (45),  $g_1(\mathbf{d}_{m,k})$ can be transformed as
\begin{eqnarray}
g_{2}(\mathbf{d})=-\mathbf{d}^{H} \mathbf{E} \mathbf{d}+\operatorname{Re}\left\{2 \boldsymbol{v}^{H} \mathbf{d}\right\}-Y,
\end{eqnarray}
where $\mathbf{E}=\operatorname{diag}\left(\mathbf{E}_1, \ldots, \mathbf{E}_M\right)$ and $\boldsymbol{v} \in \mathbb{C}^{N_{\mathrm{RF}} M K \times 1}=\left[\boldsymbol{v}_{1,1}^T, \boldsymbol{v}_{1,2}^T, \ldots, \boldsymbol{v}_{1, K}^T, \boldsymbol{v}_{2,1}^T, \boldsymbol{v}_{2,2}^T, \ldots, \boldsymbol{v}_{M, K}^T\right]^{\boldsymbol{T}}$. Moreover, we define $\widehat{\mathbf{C}}_m=\mathbf{F}_m^{H} \mathbf{F}_m$, $\mathbf{C}_m=I_K \otimes \widehat{\mathbf{C}}_m$, $\mathbf{C}=\operatorname{diag}\left(\mathbf{C}_1, \ldots, \mathbf{C}_M\right)$. Consequently, $\mathrm{P3}$ can be reformulated as
\begin{subequations}\label{OptA}
\begin{align}
\;\;\mathrm{P4}:&\min _{\mathbf{d}}\mathbf{d}^{H} \mathbf{E} \mathbf{d}-\operatorname{Re}\left\{2 \boldsymbol{v}^{H} \mathbf{d}\right\}\label{OptA0}\\
{\rm{s.t.}}\;\;&\operatorname{Tr}(\mathbf{d}^{H}\mathbf{C}\mathbf{d}) \leq P_{\text {max }}\label{OptA1}.
\end{align}
\end{subequations}
Since matrices $\mathbf{E}$ and $\mathbf{C}$ are positive semidefinite, $\mathrm{P4}$ is a standard QCQP problem, hence the optimal $\mathbf{d}$ can be calculated by existing  convex optimization algorithms or toolboxes \cite{ref30}.
\subsection{Solution of $\mathbf{A}_{i}$}
Based on given $\boldsymbol{\rho}$ and $\mathbf{d}$, the objection function in (44a) with respect to $\mathbf{A}_{i}$ is still a sum of multiple fractions, namely
\begin{eqnarray}
g_3(\mathbf{A}_{i})=\sum_{k=1}^{K} \sum_{m=1}^{M}\frac{\left(1+\rho_{m, k}\right)\left|\Psi_{i,k, m, k}(\mathbf{A}_{i})\right|^{2}}{\sum_{j=1}^{K}\left|\Psi_{i,k, m, j}(\mathbf{A}_{i})\right|^{2}+\sigma_{m, k}^{2}},
\end{eqnarray}
where $\Psi_{i,k, m, j}(\mathbf{A}_{i})=\mathbf{h}_{ m, k} \mathbf{A}_{i} \boldsymbol{\Phi}_{i,m} \mathbf{G}_{m}\mathbf{F}_{m} \mathbf{d}_{m, j}$.
Thus, MCQT is applied again to equivalently transform the objection function~as
\begin{eqnarray}
\begin{aligned}
g_4(\mathbf{A}_{i}, \boldsymbol{\epsilon})&=\sum_{\mathrm{k}=1}^K \sum_{\mathrm{m}=1}^{\mathrm{M}} 2 \sqrt{\left(1+\rho_{m, k}\right)} \operatorname{Re}\left\{\epsilon_{m, k}^{H} \Psi_{i,k, m, k}\right\}\\
&-\sum_{\mathrm{k}=1}^K \sum_{\mathrm{m}=1}^{\mathrm{M}} \epsilon_{m, k}^{H}\left(\sum_{\mathrm{j}=1}^K\left|\Psi_{i,k, m, j}\right|^2+\sigma_{m, k}^2\right) \epsilon_{m, k},
\end{aligned}
\end{eqnarray}
where $\boldsymbol{\epsilon}=\left[\epsilon_{1,1}, \epsilon_{1,2}, \ldots, \epsilon_{1, K}, \epsilon_{2,1}, \epsilon_{2,2}, \ldots, \epsilon_{M, K}\right]$ is the auxiliary variable. Then, the problem of solving $\mathbf{A}_{i}$ can be transformed~as
\begin{subequations}\label{OptA}
\begin{align}
\;\;\mathrm{P5}:&\max _{\mathbf{A}_{i}, \boldsymbol{\epsilon}}g_4(\mathbf{A}_{i}, \boldsymbol{\epsilon})\label{OptA0}\\
{\rm{s.t.}}\;\;&\sum_{i \in \{ \mathbb{R}, \mathbb{T}\}}\beta_{i,r,n}^{2}\leq 1, \forall r \in \mathcal{R}, n=1,2, \ldots, N_{\rm RIS}\label{OptA1}.
\end{align}
\end{subequations}
Given other variables, the function $g_4$ is a convex function with respect to $\epsilon_{m, k}$, $\forall m \in \left[1,2, \cdots, M\right]$, $ \forall k \in \mathcal{K}$. The optimal $\boldsymbol{\epsilon}$ is obtained by setting $\partial g_4 / \partial \epsilon_{m, k}=0$, which is calculated as
\begin{eqnarray}
\epsilon_{m, k}^{*}=\frac{\sqrt{\left(1+\rho_{m, k}\right)}\left|\Psi_{i,k, m, k}(\mathbf{A}_{i})\right|}{\sum_{j=1}^{K}\left|\Psi_{i,k, m, j}(\mathbf{A}_{i})\right|^{2}+\sigma_{m, k}^{2}}.
\end{eqnarray}
After obtaining the auxiliary variable $\boldsymbol{\epsilon}$, $\mathrm{P5}$ can be transformed~as
\begin{subequations}\label{OptA}
\begin{align}
\;\;\mathrm{P6}:&\max _{\mathbf{A}_{i}}g_4(\mathbf{A}_{i})\label{OptA0}\\
{\rm{s.t.}}\;\;&\sum_{i \in \{ \mathbb{R}, \mathbb{T}\}}\beta_{i,r,n}^{2}\leq 1, \forall r \in \mathcal{R}, n=1,2, \ldots, N_{\rm RIS}\label{OptA1}.
\end{align}
\end{subequations}
By defining
\begin{eqnarray}
\begin{aligned}
\epsilon_{m, k}^H \Psi_{i,k, m, j}\left(\mathbf{A}_{i}\right)&=\epsilon_{m, k}^H \mathbf{h}_{m, k} \mathbf{A}_i \boldsymbol{\Phi}_{i,m} \mathbf{G}_m \mathbf{F}_{m}\mathbf{d}_{m, j}\\
&=\boldsymbol{\beta}_i^{H} \operatorname{diag}\left(\epsilon_{m, k}^H \mathbf{h}_{m, k}\right) \boldsymbol{\Phi}_{i,m} \mathbf{G}_m \mathbf{F}_{m} \mathbf{d}_{m, j}\\&
=\boldsymbol{\beta}_i^{H} \mathbf{q}_{i, m, k, j},
\end{aligned}
\end{eqnarray}
the function $g_4(\mathbf{A}_{i})$ can be transformed as
\begin{equation}
\begin{split}
&g_4(\mathbf{A}_{i})=\sum_{\mathrm{k}=1}^K \sum_{\mathrm{m}=1}^{\mathrm{M}} 2 \sqrt{\left(1+\rho_{m, k}\right)} \operatorname{Re}\left\{\boldsymbol{\beta}_i^{H} \mathbf{q}_{i, m, k, k}\right\}\\
&-\sum_{\mathrm{k}=1}^K \sum_{\mathrm{m}=1}^{\mathrm{M}} \epsilon_{m, k}^{H}\left(\sum_{\mathrm{j}=1}^K\left|\boldsymbol{\beta}_i^{H} \mathbf{q}_{i, m, k, j}\right|^2 \epsilon_{m, k}\right)-\sum_{\mathrm{k}=1}^K \sum_{\mathrm{m}=1}^{\mathrm{M}} \epsilon_{m, k}^{H}\sigma_{m, k}^2 \epsilon_{m, k},
\end{split}
\end{equation}
where $\boldsymbol{\beta}_i=\left[\beta_{i, 1,1}, \ldots, \beta_{i, 1, N_{\mathrm{RIS}}}, \beta_{i, 2,1}, \ldots, \beta_{i, R, N_{\mathrm{RIS}}}\right]^{T}$ is the reflection/transmission amplitude coefficients vector, $\mathbf{q}_{i, m, k, j}=\operatorname{diag}\left(\boldsymbol{\epsilon}_{m, k}^H \mathbf{h}_{m, k}\right) \boldsymbol{\Phi}_{i, m} \mathbf{G}_m \mathbf{F}_{m} \mathbf{d}_{m, j}$. Next, the objective function $g_4(\mathbf{A}_{i})$ is further equivalent as
\begin{eqnarray}
g_4(\mathbf{A}_{i})=\sum_{i \in \{ \mathbb{R}, \mathbb{T}\}}\left(-\boldsymbol{\beta}_i^{H} \boldsymbol{\Delta}_i \boldsymbol{\beta}_i+\operatorname{Re}\left\{2 \boldsymbol{\beta}_i^{H} \upsilon\right\}\right)-\Omega,
\end{eqnarray}
where $\boldsymbol{\Delta}_i=\sum_{k \in  \mathcal{K}_{i}}\sum_{m=1}^{M} \sum_{j=1}^{K} \mathbf{q}_{i,k, m, j} \mathbf{q}_{i,k, m, j}^{H}, \forall i \in \{ \mathbb{R}, \mathbb{T}\}$, $\upsilon_i=\sum_{k \in  \mathcal{K}_{i}} \sum_{m=1}^{M} \sqrt{1+\rho_{m, k}} \mathbf{q}_{i, m, k,k}$, and $\Omega=\sum_{k=1}^{K} \sum_{m=1}^{M} \epsilon_{m, k}^{H} \sigma_{m, k}^{2} \epsilon_{m, k}$.

Consequently, $\mathrm{P6}$ can be rewritten as
\begin{subequations}\label{OptA}
\begin{align}
\;\;\mathrm{P7}:&\min _{\mathbf{A}_{i}}\sum_{i \in \{ \mathbb{R}, \mathbb{T}\}}\left(\boldsymbol{\beta}_i^{H} \boldsymbol{\Delta}_i \boldsymbol{\beta}_i-\operatorname{Re}\left\{2 \boldsymbol{\beta}_i^{H} \upsilon\right\}\right)\label{OptA0}\\
{\rm{s.t.}}\;\;&\sum_{i \in \{ \mathbb{R}, \mathbb{T}\}}\beta_{i,r,n}^{2}\leq 1, \forall r \in \mathcal{R}, n=1,2, \ldots, N_{\rm RIS}\label{OptA1}.
\end{align}
\end{subequations}
Observe that the matrix $\boldsymbol{\Delta}_i$ is positive semidefinite, hence the objective function is convex. Therefore, the problem $\mathrm{P7}$ can be solved by ADMM \cite{ref31}.

Through the above analysis, the joint optimization scheme of analog beamforming $\mathbf{F}_{\rm{A}}$, time delay $\mathbf{F}_{m}^{\rm{td}}$, digital beamforming $\mathbf{d}_{m, k}$ at the BS, time delay coefficients $\mathbf{T}_{i,m}$, first-layer phase-shift coefficients $\boldsymbol{\Phi}_{i,1}$, second-layer phase-shift coefficients $\boldsymbol{\Phi}_{i,2}$ and reflection/transmission amplitude coefficients $\mathbf{A}_{i}$ at the STAR-RISs design is summarized in $\bf Algorithm \hspace*{0.02in} 1$. To be specific, the analog beamforming and time delays of the BS can be first derived by the different STAR-RISs' physical directions, and the time delays, and double-layer phase-shift of the STAR-RISs are obtained by users' locations, which is analysed in Section II-B. Then, the auxiliary variables $\boldsymbol{\rho}$, $\boldsymbol{\varpi}$ and $\boldsymbol{\epsilon}$, the transmit digital beamforming $\mathbf{d}_{m,k}$, and the STAR-RIS amplitude coefficients $\mathbf{A}_{i}$ are iteratively updated until convergence.
\begin{algorithm}[t]
\caption{Beamforming Design for Solving $\mathrm{P1}$}
{\bf Input:}
Channels $\mathbf{h}_{r, m, k}$, $\mathbf{G}_{r, m}$.\\
{\bf Initialization:}
Digital beamforming vector $\mathbf{d}_{m, k}^{(0)}$ and the reflection/transmission amplitude coefficients $\mathbf{A}_{i}$.

Obtain the analog beamforming matrix $\mathbf{F}_{m}$ via (32) and (35).\\

Calculate the STAR-RIS reflection/transmission phase-shift coefficients matrix $\boldsymbol{\Phi}_{i,m}$ via (36)-(38).\\

{\bf while} no convergence {\bf do}\\
Update $\boldsymbol{\rho}$ via (43);\\
Update $\boldsymbol{\varpi}$ via (46);\\
Update $\mathbf{d}_{m, k}$ via solving $\mathrm{P4}$;\\
Update $\boldsymbol{\epsilon}$ via (56);\\
Update $\mathbf{A}_{i}$ via solving $\mathrm{P7}$;\\
{\bf end while}

{\bf Output:}
Analog beamforming matrix $\mathbf{F}_{m}$, phase-shift coefficients matrix $\boldsymbol{\Phi}_{i,m}$, digital beamforming vector $\mathbf{d}_{m, k}$, amplitude coefficients $\mathbf{A}_{i}$.
\end{algorithm}
\subsection{Computational Complexity analysis}
The overall computational complexity of $\bf Algorithm \hspace*{0.02in} 1$ is mainly caused by the update of the variables $\boldsymbol{\rho}$, $\boldsymbol{\varpi}$, $\mathbf{d}_{m,k}$, $\boldsymbol{\epsilon}$ and $\mathbf{A}_{i}$. In each iteration, obtaining the optimal solution of $\boldsymbol{\rho}$, $\boldsymbol{\varpi}$ and $\boldsymbol{\epsilon}$ requires $\mathcal{O}\left(K^2 N_{t} M+(K+1)KM+2KM\right)$, $\mathcal{O}\left(K^2 N_{t} M+(K+1)KM+KM\right)$ and $\mathcal{O}\left((K+1)KM+KM\right)$ respectively.
Updating the digital beamforming vector $\mathbf{d}_{m, k}$ is about $\mathcal{O}\left( K^3 N_{t}^{3} M^3\right)$. The reflection/transmission amplitude coefficients matrix $\mathbf{A}_{i}$ owns a complexity of $\mathcal{O}\left((R N_{\rm RIS})^3 N_{t}^{3}\right)$. Therefore, the overall computational complexity of $\bf Algorithm \hspace*{0.02in} 1$ can be approximated as $\mathcal{O}\left(I_o(K^3 N_{t}^{3} M^3+(R N_{\rm{RIS}})^3 N_{t}^{3})\right)$ , where $I_o$ is the number of iterations for the convergence.
\section{Numerical Results}
In this section, we present the numerical results to validate the performance of our proposed schemes. The investigated scenario is illustrated in Fig. 6, where reflection/transmission users are located on half-circles centered at $(0m, 10m, 0m)$ with a radius of $1 m$. The main system parameters are shown in Table I. The BS is equipped with $N_{\rm RF} = 4$ RF chains to serve two reflection users and two transmission users. We assume that each RF chain is connected to $K_{\rm T}=16$ TDs and each STAR-RIS composes $N_{\rm{RIS}}= N_1\times N_2=8\times 8$ elements.
\begin{figure}[htbp]
\centering
    \label{RIS_subcarrier} 
    \includegraphics[width=8.5cm,height=5.5cm]{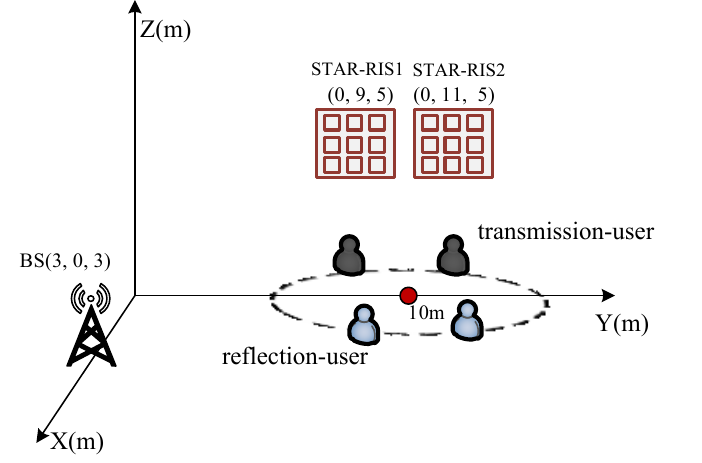}
	\caption{ The simulation scenario setup.}
\end{figure}

\begin{table}[htp]
\begin{center}
\caption{System parameters}
\label{table:1}
\begin{tabular}{|c|c|c|}
\hline   \textbf{Parameters} & \textbf{Value} \\
\hline   Central frequency & $f_{c}=100$ GHz  \\
\hline   Bandwidth & $B=10$ GHz  \\
\hline   Number of antennas & $N_t=128$  \\
\hline   Number of subcarriers & $M=8$  \\
\hline   Number of TDs &  $K_{\rm{T}}=16$ \\
\hline   Number of STAR-RISs &  $R=2$ \\
\hline   Number of STAR-RIS elements &  $N_{\rm{RIS}}=64$ \\
\hline   Number of users &  $K=4$ \\
\hline   Number of RF chains & $N_{\rm{RF}}=4$  \\
\hline   Maximum transmit power & $P_{\rm{max}}=15$ W  \\
\hline   Noise power & $\sigma_{m,k}^{2} = -85$ dBm  \\
\hline
\end{tabular}
\end{center}
\end{table}
To show the convergency of $\bf Algorithm \hspace*{0.02in} 1$, Fig.~7 plots the sum rate versus the number of iterations $I_o$, under four different structure schemes, including STAR-RIS with fully-connected structure, STAR-RIS with sub-connected structure, STAR-RIS with conventional structure, BS and STAR-RIS without TDs.  For the STAR-RIS with sub-connected structure, we divide each STAR-RIS into $S = 2 \times 2$ sub-surfaces and each sub-surface consists of $L =4 \times 4$ elements. We find that the sum rate speedily converges to stabilization with 12 iterations under all schemes, which shows the effectiveness of the proposed algorithm. Furthermore, the STAR-RIS with fully-connected structure owns the highest sum rate, while the sum rate under the BS and STAR-RIS without TDs is the lowest. However, the hardware complexity of the former is much higher than that of the latter. Therefore, to tradeoff the hardware complexity and system performance, the STAR-RIS with sub-connected structure is the best choice.
\begin{figure}[t]
\centering
    \label{RIS_subcarrier} 
    \includegraphics[width=8.5cm,height=5.5cm]{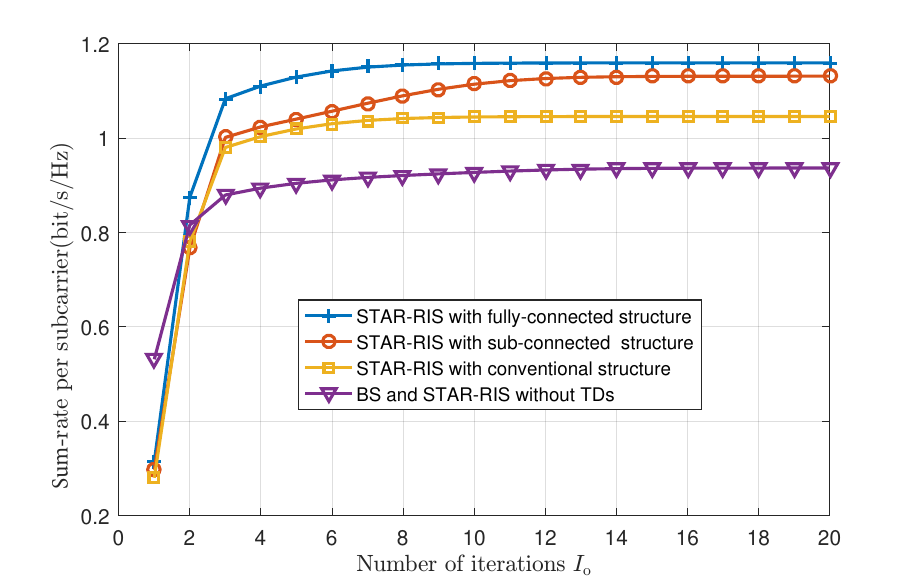}
	\caption{Sum rate versus the number of iterations $I_o$.}
\end{figure}
\begin{figure}[t]
\centering
    \label{RIS_subcarrier} 
    \includegraphics[width=8.5cm,height=5.5cm]{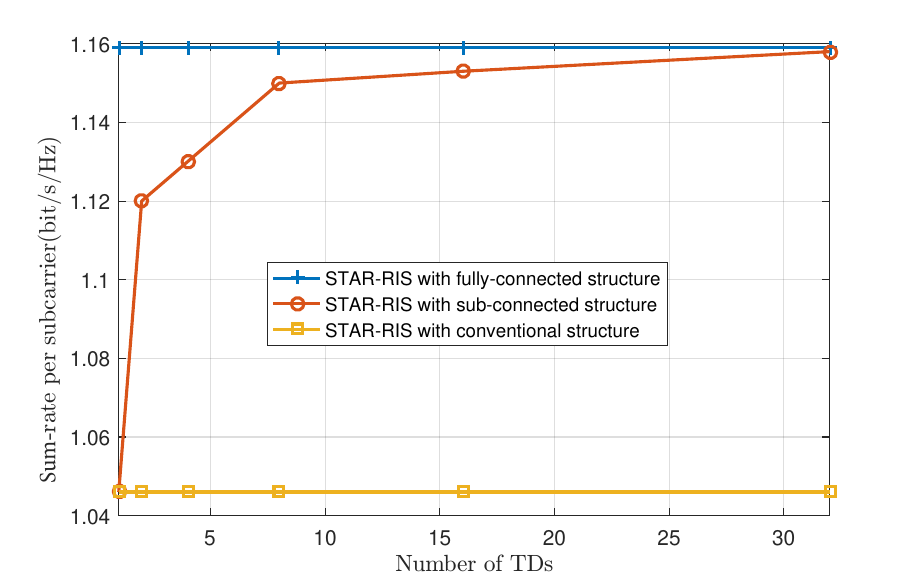}
	\caption{Sum rate versus the number of TDs.}
\end{figure}

Fig. 8 provide the sum rate versus the number of TDs, i.e., the number of sub-surfaces $S$. We can find that the sum rate under the STAR-RIS with sub-connected structure increases with the number of TDs $S$ grows up. It is noteworthy that small quantity of TDs can already obtain a good system performance. For example, it can reach more than $98\%$ of the optimal sum rate when $S \geq 16$. Meanwhile, its hardware complexity and power consumption is much lower than that of the STAR-RIS with fully-connected structure. It is obvious that the sum rate is the lowest for the STAR-RIS with conventional~structure.
\begin{figure}[t]
\centering
    \label{RIS_subcarrier} 
    \includegraphics[width=8.5cm,height=5.5cm]{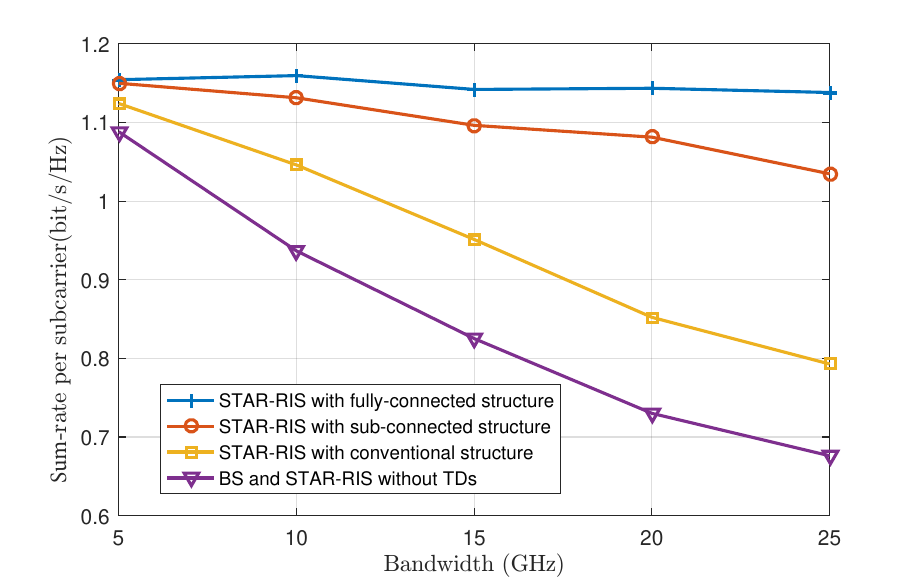}
	\caption{Sum rate versus bandwidth.}
\end{figure}

Then, Fig.~9 evaluates the sum rate versus bandwidth. The number of TDs is $S = 4$ for each STAR-RIS with the sub-connected structure. It is well know that the beam split effect increases with the bandwidth. Therefore, for the BS and STAR-RIS without TDs structure, the sum rate speedily decreases with bandwidth increases. For the STAR-RIS with conventional structure, although STAR-RIS is not equipped with TDs, the BS still apply the TDs to mitigate partial beam split effect, and thus its sum rate is higher than that under the BS and STAR-RIS without TDs. Besides, since the number of TDs  under the STAR-RIS with sub-connected structure is smaller than that of the STAR-RIS with fully-connected structure, its sum rate is a litter lower.

\begin{figure}[t]
\centering
    \label{RIS_subcarrier} 
    \includegraphics[width=8.5cm,height=5.5cm]{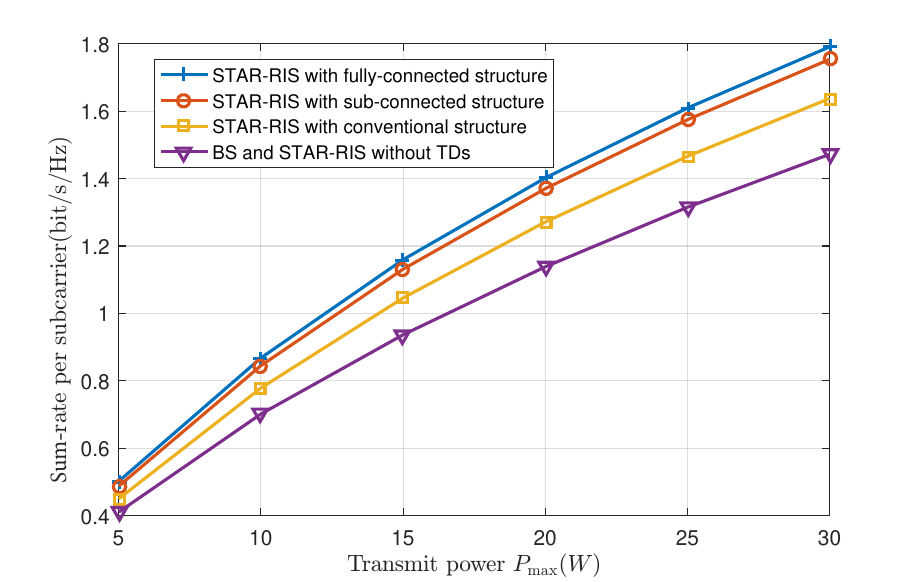}
	\caption{Sum rate versus the maximum transmit power.}
\end{figure}
\begin{figure}[t]
\centering
    \label{RIS_subcarrier} 
    \includegraphics[width=8cm,height=5cm]{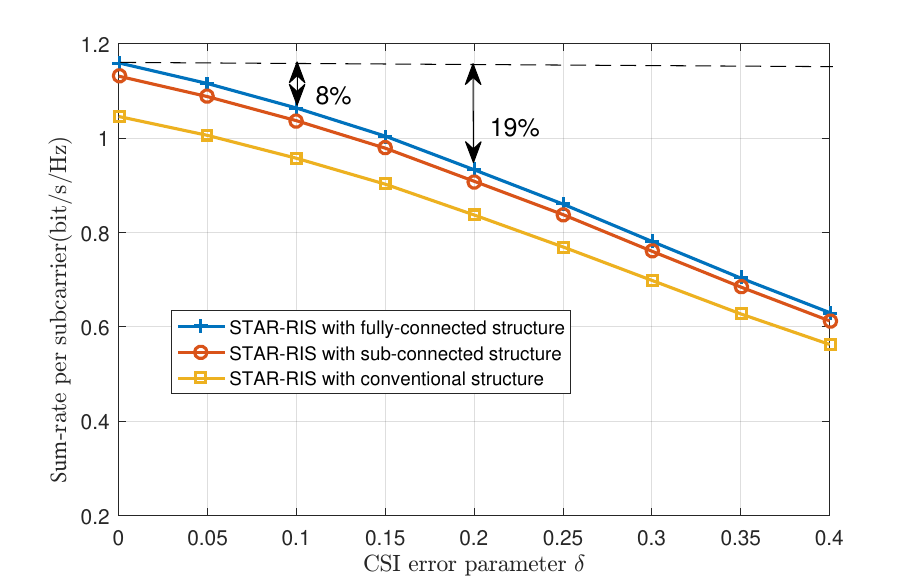}
	\caption{Sum rate versus $\delta$.}
\end{figure}
Fig. 10 plots the sum rate versus the maximum transmit power $P_{\rm max}$. The number of TDs is $S = 4$ for each STAR-RIS with the sub-connected structure. We can note that the sum rate grows up with $P_{\rm{max}}$ under different schemes. Particularly, the STAR-RIS with sub-connected structure still achieves very closed performance to the STAR-RIS with fully-connected structure. It demonstrates the effectiveness of the sub-connected structure, which can trade off the hardware cost and performance.

Additionally, in this paper, we provide the robustness of the joint beamforming against the impacts of imperfect CSI. The CSI uncertainties can be modeled as
\begin{eqnarray}
\widetilde{\hbar}=\hbar+e,
\end{eqnarray}
where $\widetilde {\hbar}$ and $\hbar$ represent estimated and real channel, respectively, $e$ indicates the independent estimation error following complex Gaussian distribution with zero mean, i.e. $e \sim \mathcal{C} \mathcal{N}\left(0, \sigma_{e}^{2}\right)$. We assume that the variance $\sigma_{e}^{2}$ , i.e. the error power, satisfies $\sigma_e^2 \triangleq \delta|\hbar|^2$ where $\delta$ characterizes the level of CSI~error.

Finally, the sum rate versus the CSI error parameter $\delta$ is illustrated in Fig. 11. We can find that the performance loss grows with the increasing of $\delta$. The main reason is that the accuracy of the estimation angles deteriorates, which causes the beam misalignment. In particular, for the STAR-RIS with fully-connected structure, compared with the perfect CSI(i.e. $\delta = 0$), the system suffers a loss of $8\%$ with the error power $\delta = 0.1$, and a loss of $19\%$ with $\delta = 0.2$. Besides, the STAR-RIS with sub-connected structure always outperforms the structure without TD and close to the STAR-RIS with fully-connected structure at any CSI error, which demonstrates the effectiveness and robustness of the proposed scheme.
\section{Conclusions}
In this paper, we investigated the STAR-RIS-assisted THz communication system and proposed two STAR-RIS structures to overcome the beam split effect. Meanwhile, we developed an joint optimization scheme of the hybrid analog/digital beamforming, time delays at the BS as well as the double-layer phase-shift coefficients, time delays and amplitude coefficients in a multiple STAR-RISs to maximize the system sum rate. Numerical results verified that the STAR-RIS with sub-connected structure can realize a better tradeoff the system performance and hardware cost. In future, we will investigate the practical phase shift model, and utilize the beam split effect to realize target sensing in the STAR-RIS-assisted wideband THz systems.

\end{document}